\pdfoutput=1

\documentclass[cernpreprint,texlive=2016,txfonts,UKenglish]{latex/atlasdoc} 

\usepackage[biblatex=false,subcaption=true]{latex/atlaspackage}
\usepackage{latex/atlasbiblatex}

\usepackage[BSM]{latex/atlasphysics}

\graphicspath{{logos/}{figures/}}

\hypersetup{pdftitle={ATLAS document},pdfauthor={The ATLAS Collaboration}}

\AtlasRefCode{SUSY-2016-23}
 
\PreprintIdNumber{CERN-PH-2017-173}
 
 \AtlasJournalRef{\EPJC 78 (2018) 154}
 \AtlasDOI{10.1140/epjc/s10052-018-5583-9}

\hypersetup{
  pdftitle={ATLAS document},
  pdfauthor={The ATLAS Collaboration},
  colorlinks=true,
  urlcolor=blue
}

\usepackage{url}

\usepackage[textwidth=0.1\textwidth,disable]{todonotes} 
\usepackage{units}
\usepackage{xspace} 
\usepackage{rotating} 
\usepackage{tabularx} 
\usepackage{cite}     

\newcommand{\hd}{\hphantom{0}} 
\newcommand{\inmath}[1]{\ensuremath{#1}\xspace}
\newcommand{\percent}[1]{{#1}\%}

\newcommand{\CLs}{CL$_\mathrm{s}$ }
\newcommand{\tttev}{\inmath{\sqrt{s} = \unit[13]{TeV}}}
\newcommand{\ourintlumi}{36.1~\ifb}

\renewcommand{\stau}{\inmath{\widetilde{\tau}}}

\newcommand*{\chinoonemp}{\ensuremath{\mathchoice%
      {\displaystyle\raise.4ex\hbox{$\displaystyle\tilde\chi^\mp_1$}}%
         {\textstyle\raise.4ex\hbox{$\textstyle\tilde\chi^\mp_1$}}%
       {\scriptstyle\raise.3ex\hbox{$\scriptstyle\tilde\chi^\mp_1$}}%
       {\scriptscriptstyle\raise.3ex\hbox{$\scriptscriptstyle\tilde\chi^\mp_1$}}}\xspace}

\newcommand{\mttwo}{\ensuremath{m_\mathrm{T2}}}
\newcommand{\ConepConem}{$\chinoonep\chinoonem$\xspace}
\newcommand{\ConepmNtwo}{$\chinoonepm\ninotwo$\xspace}

\AtlasTitle{%
  Search for the direct production of charginos and neutralinos 
  in final states with tau leptons in \tttev $pp$ collisions with the ATLAS detector
}

\author{The ATLAS Collaboration}
\date{\today}

\AtlasAbstract{%
  A search for the direct production of charginos and neutralinos
  in final states with at least  two hadronically decaying tau leptons is presented.  
  The analysis uses a dataset of $pp$ collisions corresponding to an integrated luminosity of 
  \ourintlumi, recorded with the ATLAS detector at the Large Hadron Collider
    at a centre-of-mass energy of 13~TeV.
  No significant deviation from the expected Standard Model background is observed. 
  Limits are derived in scenarios of \ConepConem\ pair production and of \ConepmNtwo\ and \ConepConem\ production in simplified models where 
the neutralinos and charginos decay solely via intermediate left-handed staus and tau sneutrinos, and 
the mass of the $\tilde{\tau}_{\mathrm L}$ state is set to be halfway between the masses of the \chinoonepm\ and the \ninoone.
  Chargino masses up to 630 GeV are excluded at \percent{95} confidence level
 in the scenario of direct production of \ConepConem\  for a massless \ninoone. 
  Common \chinoonepm and \ninotwo\ masses up to 760 GeV  are excluded 
  in the case of production of \ConepmNtwo\ and \ConepConem\ assuming a massless \ninoone.
  Exclusion limits for additional benchmark scenarios with large and small mass-splitting between the 
  \chinoonepm and the \ninoone  are also studied by varying the $\tilde{\tau}_{\mathrm L}$ mass 
  between the masses of the \chinoonepm\ and the \ninoone.
}

\begin{document}
\maketitle

\section{Introduction}

Supersymmetry (SUSY)~\cite{Golfand:1971iw,Volkov:1973ix,Wess:1974tw,Wess:1974jb,Ferrara:1974pu,Salam:1974ig,Martin:1997ns}  
postulates the existence of a superpartner, referred to as a \textit{sparticle}, 
 whose spin differs by one half unit from each corresponding Standard Model (SM) partner.
 In models that conserve $R$-parity~\cite{Farrar:1978xj},
 sparticles are always produced in pairs, and the lightest supersymmetric particle (LSP) is stable and provides a dark-matter 
 candidate~\cite{DM,Goldberg:1983nd,Ellis:1983ew}. 

In SUSY models, the sector of sparticles with only electroweak interactions contains charginos ($\susy{\chi}^{\pm}_{i}$, $i$ = 1, 2 in order of increasing masses), neutralinos 
($\susy{\chi}^{0}_{j}$, $j$ = 1, 2, 3, 4 in order of increasing masses),  sleptons (\slepton), and sneutrinos ($\susy{\nu}$). 
Charginos and neutralinos are the mass eigenstates formed from the linear superpositions of the superpartners of the
charged and neutral Higgs bosons and electroweak gauge bosons.
The sleptons are the superpartners of the leptons and  are referred to as left or right (\sleptonL or \sleptonR)
depending on the chirality of their SM partners.
The slepton mass eigenstates are a mixture of \sleptonL and \sleptonR, and are labelled as $\susy{\ell}_1$ and
$\susy{\ell}_2$ (with $\susy{\ell}_{k}$, $k$ = 1, 2 in order of increasing masses).
In this work, only the \chinoonepm, the \ninotwo, the \ninoone, and the scalar superpartner of the left-handed tau lepton (the
stau, \stauL) and of the tau neutrino (the tau sneutrino, $\susy{\nu}_{\tau_L}$) are assumed to be sufficiently light to be 
produced at the Large Hadron Collider (LHC)~\cite{LHC:2008}. 

Although experimentally challenging, final states with tau leptons originating from stau decays are of particular interest for SUSY searches.
Models with light staus can lead to a dark-matter relic density consistent with cosmological observations \cite{Vasquez:2011}, 
and light sleptons in general could play a role in the co-annihilation of neutralinos~\cite{Belanger:2004ag,King:2007vh}.
Sleptons are expected to have masses of $\mathcal{O}$(100 GeV) in gauge-mediated~\cite{Dine:1981gu,AlvarezGaume:1981wy,Nappi:1982hm,Dine:1993yw, Dine:1994vc,Dine:1995ag} 
and anomaly-mediated~\cite{Randall:1998uk,Giudice:1998xp} SUSY breaking models.  

\begin{figure}\centering
	\includegraphics[width=0.35\textwidth]{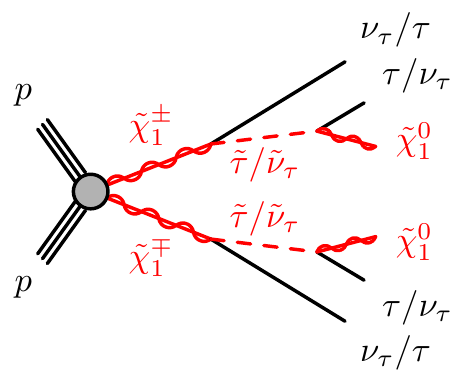}
	\includegraphics[width=0.35\textwidth]{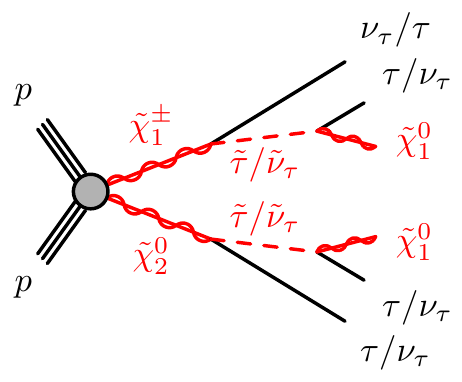}\hfil
	\caption{\label{diagrams}
		Representative diagrams for the electroweak production and decay processes of supersymmetric particles considered in this work:
		(left) \ConepConem\ and (right) \ConepmNtwo\  production.}
\end{figure}

Scenarios where the production of charginos, neutralinos, and sleptons may dominate at the LHC with respect to the production 
of squarks and gluinos can be realised in the general framework of the phenomenological Minimal Supersymmetric Standard 
Model (pMSSM)~\cite{Djouadi:1998di,Berger:2008cq}.
Two simplified models~\cite{Alwall:2008ve,Alwall:2008ag,Alves:2011wf} of 
 \ConepConem\ and \ConepmNtwo\ production are considered in this work. The models are designed to enhance the probability of experimental observation.
 In both models, the lightest neutralino is the LSP and purely bino,
the stau and tau sneutrino are assumed to be mass-degenerate, and the $\susy{\tau}_1$ is assumed to be purely \stauL.
The mass of the $\tilde{\tau}_{\mathrm L}$ state is set to be halfway between the masses of the \chinoonepm\ and the \ninoone, i.e.
$m(\tilde{\tau}_{\mathrm L}) = m(\ninoone) + x\cdot\left( m(\chinoonepm) - m(\ninoone) \right)$, with the parameter $x=0.5$. 
Other values of $x$ are also studied for selected benchmark models where $x$ is varied between $0.05$ and $0.95$ in steps of 0.1.
All sparticles other than those explicitely mentioned here are assumed to be inaccessible at the LHC energy. 
In the model characterised by \ConepmNtwo production, the \chinoonepm\ and \ninotwo\ are assumed to be pure wino and mass-degenerate.
In the model where only \ConepConem production is considered, the \chinoonepm\ is pure wino.
The above assumptions guarantee large production cross sections and short decay chains for \chinoonepm\ and \ninotwo.
Charginos and next-to-lightest neutralinos decay into the lightest neutralino via an intermediate on-shell stau or tau sneutrino, 
$\chinoonepm \rightarrow  \stau \nu_\tau (\tilde{\nu}_\tau \tau) \rightarrow \tau\nu_\tau(\nu_\tau\tau)\ninoone$,
$\ninotwo \rightarrow \stau \tau \rightarrow  \tau \tau \ninoone$, and
$\ninotwo \rightarrow \tilde{\nu}_\tau \nu_\tau \rightarrow \nu_\tau \nu_\tau \ninoone$
(see Figure~\ref{diagrams}). 

Signal events are characterised by the presence of at least two tau leptons and large missing transverse energy, $\met$, due to the
undetected neutrinos and lightest neutralinos. Final states with at least two hadronically decaying tau leptons (tau $\rightarrow$ hadrons $\nu_\tau$) are 
considered, as this choice provides the best discrimination of SUSY events of interest from SM background processes 
(mainly multi-jet, $W$+jets and diboson production). In multi-jet events passing the selection requirements described in Section \ref{sec:SR}, 
nearly all reconstructed tau leptons are misidentified jets. $W$($\rightarrow \tau \nu_\tau$)+jets events contribute due to the $\met$ from the neutrino, one tau lepton from
the $W$ decay, and one or more jets misidentified as tau leptons. The jet misidentification typically results in a mismeasurement of $\met$, which tends to assume large
values. Diboson events with $WW$ or $ZZ$ decaying into $\tau\tau\nu\nu$ final states contain two tau leptons and large $\met$ from the neutrinos.

The search described in this paper uses a dataset of $\sqrt{s}=13$~TeV $pp$ collisions collected with the ATLAS detector in 2015 and 2016,  
with an integrated luminosity of \ourintlumi.
In a previous similar search by the ATLAS Collaboration using the 8~TeV Run-1 dataset~\cite{SUSY-2013-14}, \chinoonepm masses up to 345~GeV were excluded at \percent{95} confidence level
for a massless \ninoone in the scenario of direct production of \ConepConem. 
In the case of production of \ConepmNtwo\ and \ConepConem, common \chinoonepm and \ninotwo\ masses up to 410~GeV were excluded for a massless \ninoone. 
Results of a similar search in the Run-1 dataset from the CMS Collaboration are reported in Refs.~\cite{CMS-SUS-13-006,CMS-SUS-14-021}. In Ref.~\cite{CMS-SUS-13-006}, 
charginos lighter than 320~GeV are excluded at \percent{95} confidence level in the case of a massless \ninoone. 
The combined LEP limits on the stau and chargino\footnote{
For the interval $0.1 \lesssim\Delta m(\chinoonepm,\ninoone) \lesssim 3 \GeV$, the chargino mass limit set by LEP degrades to $91.9$ $\GeV$.}
masses are $m_{\stau}>87$--$93\GeV$ (depending on $m_{\ninoone}$) and
$m_{\chinoonepm}>103.5\GeV$~\cite{LEPSUSYWG:01-03.1,Heister:2003zk,Abdallah:2003xe,Acciarri:1999km,Abbiendi:2003sc}, respectively.

\section{ATLAS detector}
\label{sec:atlas}

The ATLAS detector~\cite{PERF-2007-01} is a multi-purpose particle physics detector with 
forward-backward symmetric cylindrical geometry and nearly $4\pi$
coverage in solid angle.\footnote{
	ATLAS uses a right-handed coordinate system with its origin at the nominal 
	interaction point (IP) in the centre of the detector, and the $z$-axis along the 
	beam line. The $x$-axis points from the IP to the centre of the LHC ring, 
	and the $y$-axis points upwards. Cylindrical coordinates $(r,\phi)$ are used 
	in the transverse plane, $\phi$ being the azimuthal angle around the $z$-axis. 
	Observables labelled \textit{transverse} refer to the projection into the $x$--$y$ plane. 
	The pseudorapidity is defined in terms of the polar angle $\theta$ by $\eta=-\ln\tan(\theta/2)$. 
	}
It features an inner tracking detector (ID) surrounded by a 2 T superconducting solenoid, 
electromagnetic and hadronic calorimeters, and a muon spectrometer (MS).  The ID covers the pseudorapidity region  $|\eta| <2.5$ and consists of a 
silicon pixel detector, a silicon microstrip detector, and a transition
radiation tracker. One significant upgrade for the \tttev running period is the presence of the
insertable B-Layer~\cite{CERN-LHCC-2010-013}, an additional pixel layer close to the interaction point  which
provides high-resolution hits at small radius to improve the tracking and vertex reconstruction performance.
The calorimeters are composed of high-granularity liquid-argon (LAr) electromagnetic calorimeters with lead, copper, 
or tungsten absorbers (in the pseudorapidity region $|\eta| < 3.2$) and a steel--scintillator hadronic calorimeter (for
$|\eta|<1.7$). 
The end-cap and forward regions, spanning
$1.5<|\eta|<4.9$, are instrumented with LAr calorimeters
for both the electromagnetic and hadronic measurements. The MS surrounds the calorimeters and consists 
of three large superconducting air-core toroidal magnets, each with eight coils, a system of precision tracking chambers ($|\eta|<2.7$), and detectors 
for triggering ($|\eta|<2.4$).
A two-level trigger system is used to record events~\cite{Aaboud:2016leb}.

\section{Data and simulated event samples}
\label{sec:data}

The analysed dataset, after the application of beam, detector, and data quality
requirements,  corresponds to an integrated luminosity of \ourintlumi of $pp$ collision data
recorded in 2015 and 2016 at \tttev.
The uncertainty in the combined 2015+2016 integrated luminosity is \percent{3.2}. It is derived, following a methodology similar to that detailed in Ref.~\cite{DAPR-2013-01}, from a preliminary calibration of the luminosity scale using $x-y$ beam-separation scans performed in August 2015 and May 2016. 

Monte Carlo (MC) simulated event samples are used to estimate the SUSY signal yields
and to aid in evaluating the SM backgrounds.
Generated SM events are processed through a detailed detector simulation \cite{SOFT-2010-01} based on {\GEANT4} \cite{geant4},
 whereas SUSY events are passed 
through a fast detector simulation based on a parameterisation of the performance of the ATLAS electromagnetic and 
hadronic calorimeters~\cite{ATL-PHYS-PUB-2010-013} and {\GEANT4} elsewhere. 
All simulated events are overlaid with multiple $pp$ collisions (pile-up) simulated with the soft strong interaction processes of
\PYTHIA 8.186~\cite{Sjostrand:2007gs} using the A2 set of tuned parameters~\cite{ATL-PHYS-PUB-2012-003} and the MSTW2008LO~\cite{Martin:2009iq} PDF set. 
The simulated events are reconstructed using the same algorithms as the data, and are reweighted so that the distribution of the expected number of 
collisions per bunch crossing matches the one in the data.

\subsection{Simulated background samples}
\label{sec:SM}
 
Events with $Z/\gamma^*\to\ell\ell$ $(\ell=e,\mu,\tau)$ and $W\to\ell\nu$ produced with accompanying jets (including light and heavy flavours) were generated at 
next-to-leading order (NLO) in the strong coupling constant with \SHERPA 2.2.0 and 2.2.1~\cite{Gleisberg:2008ta,Hoeche:2012yf}.
Matrix elements (ME) were calculated for up to two additional partons at NLO and four additional partons at leading order (LO), using the Comix~\cite{Gleisberg:2008fv} and OpenLoops~\cite{Cascioli:2011va} generators and merged with the \SHERPA parton shower (PS)~\cite{Schumann:2007mg} using the ME+PS@NLO prescription~\cite{Hoeche:2012yf}.
The NNPDF3.0NNLO~\cite{Ball:2014uwa} parton distribution function (PDF) set was used in conjunction with a dedicated parton-shower tuning developed by the \SHERPA authors.
The $W/Z$+jets events were normalised using their next-to-next-to-leading order (NNLO) cross sections~\cite{Catani:2009sm}.
For \SHERPA 2.2.0 samples, a simplified scale setting prescription was used in the multi-parton matrix elements, to improve the event generation speed. 

The fully leptonic diboson processes ($VV=WW/WZ/ZZ$) were generated using \SHERPA 2.2.1
including final states with all possible combinations of charged leptons and neutrinos.
The matrix elements contain all diagrams with four electroweak vertices, and they were calculated for up to one parton ($4 \ell$, $2 \ell$+$ 2\nu$, $ZZ$, $WW$) or
no additional parton ($3 \ell$+$1 \nu$, $1 \ell$+$3 \nu$, $WZ$) at NLO and up to three partons at LO.
The NNPDF3.0NNLO PDF set was used in conjunction with a dedicated PS tuning developed by the \SHERPA authors. 
Diboson processes with one of the bosons decaying hadronically and the other leptonically were simulated using the \SHERPA 2.1.1 event generator. The matrix elements are calculated
for up to one ($ZZ$) or no ($WW$, $WZ$) additional partons at NLO and up to three additional partons at LO. The CT10~\cite{CT10} PDF set was used in conjunction
with a dedicated PS tuning developed by the \SHERPA authors.
Each of the diboson processes was normalised using the corresponding NLO cross section~\cite{ATL-PHYS-PUB-2016-002}.

The production of top-quark pairs and single top quarks in the $Wt$ and $s$-channels was performed with \POWHEG-\textsc{Box} 2~\cite{Alioli:2010xd}, with the CT10 PDF set 
in the ME calculations.
Electroweak $t$-channel single-top-quark events were generated using the \POWHEG-\textsc{Box} 1 event generator.
The PS, fragmentation, and the underlying event were simulated using \PYTHIA 6.428~\cite{Sjostrand:2006za} with the CTEQ6L1 PDF set and a corresponding set of tuned parameters called the Perugia 2012 tune~\cite{Skands:2010ak}. The EvtGen 1.2.0 program~\cite{EvtGen} was used for properties of the bottom and charm hadron decays. 
The top-quark mass was set to 172.5 GeV. The overall cross section was 
computed at NNLO in $\alpha_\mathrm{s}$, including resummation of next-to-next-to-leading-logarithm (NNLL) soft gluon terms~\cite{Czakon:2011xx} for \ttbar, to NLO+NNLL accuracy for single-top-quark
  $Wt$-channel~\cite{Kidonakis:2010ux}, and to NLO for the $t$- and $s$-channels~\cite{Kant:2014oha}.
Top-quark pair production with an additional $W$ or $Z$ boson was performed using \MGMCatNLO  2.2.2~\cite{Alwall:2014hca}, while fragmentation and hadronisation 
were simulated with \PYTHIA 8.186.  The ATLAS underlying-event tune A14~\cite{ATL-PHYS-PUB-2014-021} was used with the 
NNPDF2.3LO~\cite{Ball:2012cx} PDF set, and the cross sections were normalised using NLO~\cite{Lazopoulos:2008de,Campbell:2012dh}. 

\subsection{Simulated signal samples}
\label{sec:SUSY}

Simulated signal samples were generated using \MGMCatNLO 2.2.3 interfaced to \PYTHIA 8.186 with the A14 tune for the PS modelling, 
hadronisation, and underlying event. The ME calculation is performed at tree level and includes the emission of up to two additional partons. 
The PDF set used for the generation is NNPDF2.3LO. The ME--PS matching used the CKKW-L~\cite{CKKW} prescription, with a matching scale set to one quarter of the mass of the pair of produced particles. 
Signal cross sections were calculated to next-to-leading order in the strong
coupling constant, adding the resummation of soft gluon emission at
next-to-leading-logarithm accuracy (NLO+NLL)~\cite{Fuks:2012qx,Fuks:2013vua}. 
The nominal cross section and the uncertainty were taken from an envelope of cross-section predictions 
using different PDF sets and factorisation and renormalisation scales, following the procedure described in Ref.~\cite{Borschensky:2014cia}.

Two simplified models characterised by \ConepConem and \ConepmNtwo production are considered. 
The neutralinos and charginos decay via intermediate staus and tau sneutrinos. 
In both models, the \chinoonepm\ mass is varied between 100~GeV and 1.1~TeV in steps of 50 (100) GeV for \chinoonepm\ masses smaller (larger) than 700 GeV.
The \ninoone\ mass is varied between zero and 500~GeV with a variable spacing of 25 (50) GeV for \chinoonepm\ and \ninoone\ masses smaller (larger) than
 700 and 250 GeV respectively. A total of 159 models was generated. The parameter $x$ is fixed to 0.5.
The cross section for \ConepmNtwo (\ConepConem) production ranges from 23 (11.6) pb for a \chinoonepm\ mass of 100~GeV
  to 0.74 (0.34) fb for a \chinoonepm\ mass of 1.1~TeV. 

Two reference points are used throughout this paper to illustrate the typical features of the
 SUSY models to which this analysis is sensitive:

\begin{itemize}
 \item Reference point 1: simplified model for \ConepmNtwo production with the masses of the \chinoonepm\ and the \ninotwo equal to 600 GeV, and a massless \ninoone;
 \item Reference point 2: simplified model for \ConepConem production with the mass of the \chinoonepm\ equal to 600 GeV, and a massless
 \ninoone.
\end{itemize}

The dependence on the parameter $x$ is evaluated in two additional scenarios for both 
\ConepConem and \ConepmNtwo production where $x$ is varied between $0.05$ and $0.95$ in steps of $0.1$.
The first benchmark model has a large mass-splitting between the $\chinoonepm$ and the $\ninoone$, with
$m(\chinoonepm)$ = 600~GeV and massless \ninoone, while the second model is more compressed with 
$m(\chinoonepm)$ = 250~GeV and $m(\ninoone)$ = 100~GeV.

\section{Event reconstruction}
\label{sec:object} 

\newcommand{\pTvec}{\mathbf{p}_\mathrm{T}}
\newcommand{\qTvec}{\mathbf{q}_\mathrm{T}}
\newcommand{\qT}{q_\mathrm{T}}
\newcommand{\pTell}[1]{\mathbf{p}_\mathrm{T}^{\ell#1}}
\newcommand{\pTmiss}{\mathbf{p}_\mathrm{T}^\mathrm{miss}}

Events with at least one reconstructed primary vertex~\cite{ATL-PHYS-PUB-2015-026} are selected. A primary vertex must have 
at least two associated charged-particle tracks with transverse momentum $\pt>$~400~MeV and be consistent with the beam spot envelope. 
If there are multiple primary vertices in an event, the one with the largest $\sum\pt^2$ of the associated
tracks is chosen.

Jets are reconstructed from three-dimensional calorimeter energy clusters~\cite{PERF-2014-07} using the anti-$k_t$ algorithm~\cite{anti-kt,Cacciari:2011ma} 
with a radius parameter of 0.4. Jet energies are corrected for detector inhomogeneities, 
the non-compensating response of the calorimeter, and the impact of pile-up, using factors derived from test beam 
and $pp$ collision data, and from a detailed {\GEANT4} detector simulation \cite{PERF-2011-03, ATL-PHYS-PUB-2015-015}.
The impact of pile-up is accounted for using a technique, based on jet areas, that provides an 
event-by-event and jet-by-jet correction \cite{jetarea}.
Jets that are likely to have originated from pile-up are not considered~\cite{ATLAS-CONF-2014-018}.
Jets are required to have $\pt>20$~GeV and $\vert\eta\vert < 2.8$.
Events containing jets that are likely to have arisen from detector noise or cosmic rays are removed. 

Jets containing $b$-hadrons (\bjets) are identified using the MV2c10 algorithm, a multivariate discriminant making use of track impact parameters and
reconstructed secondary vertices~\cite{PERF-2012-04}.  Candidate \bjets are required to have $\pt > 20$~GeV and $|\eta| < 2.5$.  A working point with an
average $b$-tagging efficiency of \percent{77} for simulated \ttbar events is used~\cite{ATL-PHYS-PUB-2015-022,ATL-PHYS-PUB-2016-012}.  
The expected rejection factors for light-quark and gluon jets, $c$-quark jets, and hadronically decaying tau leptons are approximately 134, 6, and 55, respectively. 

Electron candidates are reconstructed by matching clusters in the electromagnetic calorimeter 
with charged-particle tracks in the inner detector. Electrons are required to have $\pt>10$~GeV, $|\eta|<2.47$, and to satisfy the 
`loose' working point according to a likelihood-based identification \cite{ATLAS-CONF-2016-024}.
Muon candidates are reconstructed from MS tracks matching
ID tracks. 
Muons are required to have $\pt > 10$ GeV and $\vert\eta\vert < 2.7$ and fulfil the `medium' quality criteria of Ref.~\cite{PERF-2015-10}.  
Events containing a muon candidate with a poorly measured charge-to-momentum ratio ($\sigma(q/p)\,/\,|q/p|>0.2$) are rejected.
Events are required not to contain any candidate muon with large impact parameter ($|z_0|>\unit[1]{mm}$ or $|d_0|>\unit[0.2]{mm}$),
as it may originate from cosmic rays.
The efficiencies for electrons and muons to satisfy the reconstruction, identification, and isolation criteria are measured in samples of leptonic $Z$ and $J/\psi$ decays, and corrections are applied to the simulated samples to reproduce the efficiencies in data. 

The reconstruction of hadronically decaying tau leptons is based on information from tracking in the ID and three-dimensional clusters in the electromagnetic and hadronic calorimeters.
The tau reconstruction algorithm is seeded by jets reconstructed as described above but with $p_{\mathrm T}>10\GeV$ and $|\eta|<2.5$. 
The reconstructed energies of the hadronically decaying tau candidates are corrected to the tau energy scale, which is calibrated based on simulation and in-situ measurements using $Z \rightarrow \tau \tau$ decays. 
Tau neutrinos from the tau lepton decay are not taken into account in the reconstruction and calibration of the tau energy and momentum.
Hadronic tau decay candidates are 
required to have one or three associated charged-particle tracks (prongs) and the total electric charge of those tracks must be $\pm 1$ times the electron charge.
To improve the discrimination between hadronically decaying tau leptons and jets, electrons, or muons, multivariate algorithms are used
\cite{ATL-PHYS-PUB-2015-045}. The tau identification algorithm is based on a boosted decision tree (BDT) method.
The BDT algorithms use various track and cluster variables as input to discriminate tau leptons from jets. 
For 1-prong (3-prong) tau candidates, the signal efficiencies are \percent{60} (\percent{50}), \percent{55} (\percent{40}), and \percent{45} (\percent{30}) 
for the `loose', `medium', and `tight' working points, respectively. 
In the following, tau candidates are required to satisfy the medium identification criteria for jet discrimination (`medium' tau candidates), unless otherwise stated.
For electron discrimination, an overlap-based veto is used for 1-prong tau candidates.
This requirement has about \percent{95} efficiency, and a rejection factor from 10 to 50 depending on the $\eta$ range.
Tau candidates are required to have $\pt > 20\GeV$ and $\vert\eta\vert < 2.47$, excluding the transition region between the barrel and end-cap calorimeters ($1.37 < |\eta| < 1.52$).

The simulation is corrected for differences in the efficiencies of the tau identification at both trigger and reconstruction level between data and simulation. 
For hadronically decaying tau leptons originating from prompt gauge boson decays, the corrections are calculated with a \textit{tag-and-probe} method in a sample of 
$Z \rightarrow \tau \tau$ events where one tau lepton decays hadronically and the other leptonically into a muon and two neutrinos~\cite{PERF-2013-06}. 

The measurement of the missing transverse momentum vector, $\pTmiss$, and its magnitude, $\met$, is based on the negative vectorial sum of the 
$\mathbf{\pT}$ of all identified jets, tau candidates, electrons, photons, muons, and an additional soft term. The soft term is constructed from all high-quality tracks 
that are associated with the primary vertex but not with any identified particle or jet. In this way, the missing transverse momentum is adjusted for the best 
calibration of the jets and the other identified particles, while maintaining pile-up independence in the soft term
~\cite{ATL-PHYS-PUB-2015-023,ATL-PHYS-PUB-2015-027}. 

With the reconstruction methods described above, it is possible that the same observables (tracks, calorimetric clusters) are assigned to several objects. This possible double counting of reconstructed objects is resolved in the following order.
Tau candidates close to electron or muon candidates ($\Delta R < 0.2$, where $\Delta R = \sqrt{(\Delta y)^2 + (\Delta\phi)^2}~$) are
removed, as are electrons that share a track with a muon.  For electrons close to a
jet ($\Delta R < 0.4$), the electron is removed, except when $\Delta R < 0.2$ and the
jet is not $b$-tagged, in which case the jet is removed.  
Any remaining jet within $\Delta R = 0.4$ of a muon or tau candidate is removed.

\section{Event selection}
\label{sec:SR}

The events used in this analysis passed either an \textit{asymmetric di-tau} trigger or a combined \textit{di-tau + \met} trigger.  
The asymmetric di-tau trigger requires the identification of two hadronically decaying tau candidates with $p_{\mathrm{T}, \tau_1} >$ 85 GeV and 
$p_{\mathrm{T}, \tau_2} >$ 50 GeV at trigger level for the leading and next-to-leading tau candidates respectively.
Two tau candidates with $p_{\mathrm{T}, \tau_1} >$ 35 GeV and $p_{\mathrm{T}, \tau_2} >$ 25 GeV at trigger level, and \met\ $>$  50 GeV  (at uncalibrated electromagnetic scale) are required by the di-tau + \met trigger.
In events selected by the di-tau + \met trigger, the reconstructed \met must be larger than 150 GeV. 
The trigger efficiency for correctly identified tau leptons is $\sim$\percent{80} for events where, at reconstruction level, the 
leading tau candidate has \pt $>$ 95 (50) GeV, and the next-to-leading tau candidate has \pt $>$ 65 (40) GeV for the asymmetric di-tau (di-tau + \met) trigger. 

Events are required to have at least two tau candidates with opposite electric charge. 
The reconstructed mass of any opposite-sign (OS) tau pair must be larger than 12 GeV to remove tau leptons originating from decays of low-mass resonances.
This requirement has negligible effect on the signal efficiency.
Two of the reconstructed tau candidates must 
satisfy the $\pt$ requirements to be in the region where the trigger efficiency is constant (see Table~\ref{tab:SR-def}).

To further discriminate the SUSY signal events from SM background processes, additional requirements are applied to define 
the signal region (SR) selections. 
To reject events from SM processes containing a top quark, selected events must not contain any $b$-tagged jet (\textit{$b$-jet veto}).
To suppress SM backgrounds with a $Z$ boson, events are selected by requiring that the reconstructed mass
of all oppositely charged tau pairs, $m(\tau_1, \tau_2)$, must not be within 10 GeV of the mean visible $Z$ boson mass\footnote{ 
The mean visible $Z$ boson mass is defined as the mean value of a Gaussian fit of the reconstructed mass distribution of OS 
tau pairs in a MC sample of $Z \rightarrow \tau \tau$ events with associated jets.} (79 GeV). This requirement is referred to as the \textit{$Z$-veto}.
An upper bound on the \textit{stransverse} mass \mttwo~\cite{Lester:1999tx,Barr:2003rg} is imposed to reduce contributions from \ttbar\ and $WW$ events.
The \mttwo variable is defined as:
\[
\mttwo = \min_{\qTvec}\left[\max\left(m_{{\mathrm T}, \tau_1}(\mathbf{p}_{\mathrm{T}, \tau_1},\qTvec),m_{{\mathrm T}, \tau_2}(\mathbf{p}_{\mathrm{T}, \tau_2},\pTmiss-\qTvec)\right)\right],
\]
where $\mathbf{p}_{\mathrm{T}, \tau_1}$ and $\mathbf{p}_{\mathrm{T}, \tau_2}$ are the transverse momenta of the two tau candidates,
and $\qTvec$ is the transverse momentum vector that minimises the larger of the
two transverse masses $m_{{\mathrm T}, \tau_1}$ and $m_{{\mathrm T}, \tau_2}$.
The latter masses are defined by
\[
m_{{\mathrm T}}(\pTvec,\qTvec) = \sqrt{2(\pT\qT-\pTvec\cdot\qTvec)}.
\]
In events where more than two tau candidates are selected,  \mttwo\ is computed among all possible tau pairs and the combination leading to the largest value is chosen.
For \ttbar\ and $WW$ events, in which two $W$ bosons decay leptonically and $\pTmiss$
is the sum of the transverse momenta of the two neutrinos, the \mttwo\ distribution has a kinematic end-point at the $W$ mass.
For large mass differences between the next-to-lightest neutralinos, the charginos, or the staus and the lightest neutralino, the \mttwo\
distribution for signal events extends significantly beyond this end-point.

Two SRs based on large \mttwo\ and \met requirements are defined.
SR-lowMass (SR-highMass) is designed to cover signal models where the mass difference between the \chinoonepm\ and \ninoone\ 
is smaller (larger) than 200 GeV. In SR-lowMass, only the di-tau+\met trigger is used. This trigger has high efficiency in
selecting events with tau leptons originating from \chinoonepm\ and \ninotwo\  
decays in models where the mass difference between the parent particle and the
\ninoone\ is small. The main discriminating requirement is  $\mttwo>$ 70 GeV.

In SR-highMass, events are selected by the di-tau+\met trigger or by the asymmetric di-tau trigger. 
If the event is selected by the di-tau+\MET trigger, the leading tau candidate 
threshold is raised to ${p}_{\mathrm{T}, \tau_1}>$ 80 GeV.
If the event is selected by the asymmetric di-tau trigger, $\met >$ 110 GeV is required.
At least one of the tau candidates must satisfy the tight identification criteria for jet discrimination (`tight' tau candidate).
In addition, the two leading tau candidates must satisfy $m(\tau_1, \tau_2) > 110$ GeV
and $\mttwo >$ 90 GeV.
The requirements for both SRs are summarised in Table~\ref{tab:SR-def}. 
The two SRs are not mutually exclusive.

\begin{table}[!hbp]
\caption{Signal region definitions.}
\begin{center}
\begin{tabular}{|c|c|c|}
\hline 
SR-lowMass   &\multicolumn{2}{c|}{SR-highMass }                \\
\hline   
\hline   
\multicolumn{3}{|c|}{At least one opposite-sign tau pair $\hd\hd\hd\hd\hd\hd\hd\hd\hd\hd$}\\
\multicolumn{3}{|c|}{$b$-jet veto $\hd\hd\hd\hd\hd\hd\hd\hd\hd\hd\hd\hd$}\\
\multicolumn{3}{|c|}{$Z$-veto $\hd\hd\hd\hd\hd\hd\hd\hd\hd\hd\hd\hd$} \\
At least two medium tau candidates    & \multicolumn{2}{c|}{at least one medium and one tight tau candidates}  \\
--  & \multicolumn{2}{c|}{$m(\tau_1, \tau_2) >$ 110 GeV}  \\
$\mttwo>$ 70 GeV &  \multicolumn{2}{c|}{$\mttwo>$ 90 GeV} \\
Di-tau+\met trigger & di-tau+\met trigger & asymmetric di-tau trigger   \\ 
$\met >$ 150 GeV & $\met >$ 150 GeV    & $\met >$ 110 GeV           \\
                                       
${p}_{\mathrm{T}, \tau_1}>$ 50 GeV & ${p}_{\mathrm{T}, \tau_1}>$ 80 GeV & ${p}_{\mathrm{T}, \tau_1}>$ 95 GeV\\
${p}_{\mathrm{T}, \tau_2}>$ 40 GeV & ${p}_{\mathrm{T}, \tau_2}>$ 40 GeV & ${p}_{\mathrm{T}, \tau_2}>$ 65 GeV\\
\hline                              
\end{tabular}
\end{center}
\label{tab:SR-def}
\end{table}

\section{Standard Model background estimation}\label{sec:backgrounds}

The main SM processes contributing to the selected final states are
multi-jet, $W$+jets and diboson production.
Background events may contain a combination of `real' tau leptons, defined as
correctly identified prompt tau leptons, or `fake' tau leptons, which can originate from a 
misidentified light-flavour quark or gluon jet, an electron, or a muon. 

In multi-jet events nearly all tau candidates are misidentified jets. 
The multi-jet contribution in the SRs is estimated from data, as described in Section
\ref{sec:BG_QCD}.
The contribution arising from heavy-flavour multi-jet events containing
a real tau lepton from the heavy-flavour quark decay is included in the
multi-jet estimate. 
The contribution of $W$+jets events, which contain one real tau lepton from
the $W$ decay and one or more misidentified jets, is estimated from MC
simulation and normalised to data in a dedicated control region (CR), as described in Section \ref{sec:BG_W}.

Diboson production contributes mainly with events containing real tau leptons originating from
$WW$ and $ZZ$ decaying into a $\tau\tau\nu\nu$ final state.
Additional SM backgrounds arise from $Z$+jets production, or events
that contain a top quark or a top-quark pair in association with jets or
additional $W$ or $Z$ bosons (collectively referred to as \textit{top}  background in the following). 
The contribution from real tau leptons exceeds \percent{90} in $Z$+jets and diboson production, and ranges from \percent{45} to \percent{75} in backgrounds containing
top quarks according to the MC simulation. The contribution of fake tau leptons
from heavy-flavour decays in jets is found to be negligible in MC
simulation.
To estimate the irreducible background, which includes diboson, $Z$+jets and top events, only MC simulated samples are used, as
described in Section \ref{sec:BG_irre}. 

The sources of systematic uncertainty in the background estimates are described in Section \ref{sec:Systematics}.
For each signal region a simultaneous fit based on the profile likelihood method \cite{statforumlimitsl} is performed to normalise
the multi-jet and $W$+jets background estimates and propagate systematic uncertainties, as described in Section \ref{sec:fit}.

\subsection{Multi-jet background estimation}
\label{sec:BG_QCD}

\begin{figure}[!htb]
\centering
\includegraphics[width=0.8\textwidth]{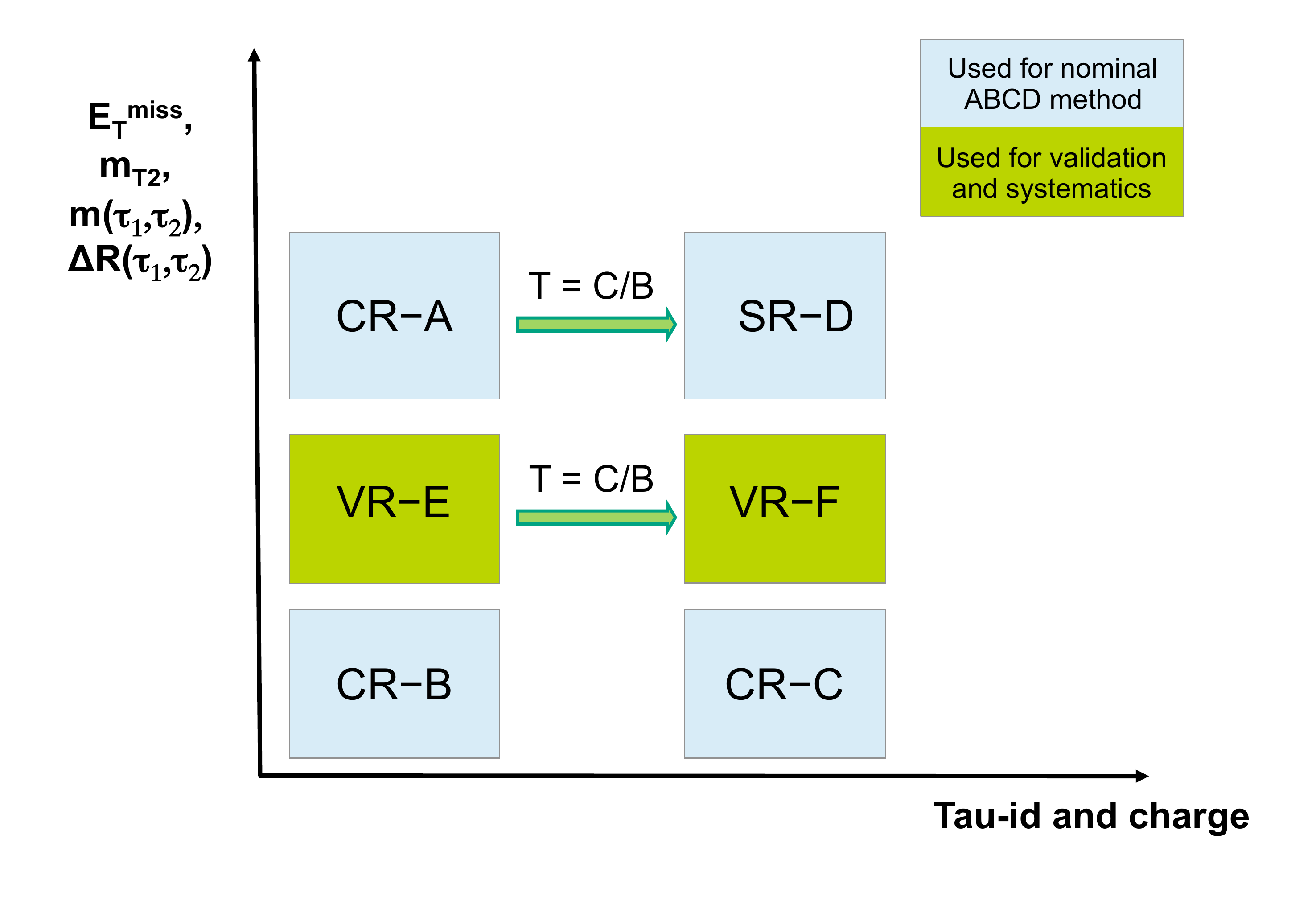}
\caption{ Illustration of the ABCD method for the multi-jet background determination.
The control regions A, B, C, and signal region D for the ABCD method described in the text 
(labelled as CR-A, CR-B, CR-C and SR-D) are drawn as light blue boxes. 
Shown in green and labelled as VR are the regions E and F, 
which are used to validate the ABCD method and to estimate the
systematic uncertainty. The definition of all regions used in the ABCD method can be found in 
Table~\ref{tab:QCD-CR}.}
\label{fig:abcd}
\end{figure}

One of the dominant backgrounds in the SRs originates from jets
misidentified as tau leptons in multi-jet production. It accounts for \percent{35} (\percent{31}) of the
total SM contribution in SR-highMass (SR-lowMass). 
This contribution is estimated from data using the so-called \textit{ABCD} method. 
All regions used for the ABCD method are schematically drawn in Figure \ref{fig:abcd}. 
Four exclusive regions, labelled as A, B, C, and D, are defined in a two-dimensional plane 
as a function of two (or more) discriminating variables that are
assumed to be uncorrelated. 
The ratio of events in the regions C and B is then equal to that in the regions D and A.
The number of events in region D, $N_{\mathrm D}$, can therefore be calculated from that in  region A, $N_{\mathrm A}$,
multiplied by the transfer factor $\mathrm{T}=N_{\mathrm C}/N_{\mathrm B}$. 
The region D corresponds to one of the SRs defined in Section \ref{sec:SR} (SR-lowMass or SR-highMass), 
whereas the regions A, B, and C are control regions defined accordingly. 
In the following, the regions A, B, C, D are labelled as CR-A, CR-B,
CR-C, and SR-D. The definition of the regions used in the ABCD method for
the multi-jet estimation is given in Table~\ref{tab:QCD-CR}.

The tau identification criteria (loose, medium or tight as defined in Section~\ref{sec:object}), the sign of the electric charges of the two taus
 (OS or same sign, SS), $m(\tau_1,\tau_2)$, $\Delta R(\tau_1,\tau_2)$, \mttwo, and \met\ are used to define CR-A, CR-B, and 
CR-C. Furthermore, two sets of validation regions (VR), VR-E and VR-F,  are defined corresponding to each SR. 
The validation regions are used to verify the extrapolation of the ABCD estimation to the SRs and to estimate the systematic 
uncertainty from the residual correlation between the tau identification and charge
requirements, and the kinematic variables \mttwo\ and \met. 

In all validation regions and both sets of CR-B and CR-C, the events passed
a \textit{di-tau} trigger instead of the di-tau + \met trigger,  due to the low \met requirements. 
The di-tau trigger requires the identification of two hadronically decaying tau candidates with transverse momenta exceeding the same 
set of thresholds as described in Section \ref{sec:SR} for the di-tau + \met trigger. The di-tau trigger was prescaled during all 2016 data-taking.

\def \hfillx {\hspace*{-\textwidth} \hfill}

\begin{table}[t]
\caption{Definition of the regions used in the ABCD method for the multi-jet estimation in SR-lowMass (left) and SR-highMass 
(right). Only those requirements that are different in the CRs/VRs with respect to the SRs are listed.}
\smallskip
\centering
\scriptsize

\begin{minipage}{0.5\textwidth}
\smallskip

\begin{tabular}{|c|c|}\hline
CR-A      & SR-D (SR-lowMass) \\
\hline
\hline
\multicolumn{2}{|c|}{Di-tau+\met\ trigger}  \\
$\ge$ 2 loose tau leptons (SS)       &   $\ge$ 2 medium tau leptons (OS)    \\
 $m(\tau_1,\tau_2) <$ 250 GeV & 	--		 	       \\
$\Delta R(\tau_1,\tau_2) >$ 1.5   &	--		 	 \\
$\met >$ 150 GeV	      &  $\met >$ 150 GeV	 \\ 
$\mttwo >$ 70 GeV	      & $\mttwo >$ 70 GeV	 \\ 
\hline
\hline
VR-E    & VR-F  \\
\hline  						    \hline
\multicolumn{2}{|c|}{Di-tau trigger}  \\
$\ge$ 2 loose tau leptons (SS)       &   $\ge$ 2 medium tau leptons (OS)    \\
 $m(\tau_1,\tau_2) <$ 250 GeV & 	--		   	 \\
$\Delta R(\tau_1,\tau_2) >$ 1.5   &	--		   	   \\
$\met >$ 40 GeV 	      &  $\met >$ 40 GeV	   \\ 
50 $<\mttwo <$~70 GeV	      & 50 $<\mttwo <$~70 GeV	   \\ 
\hline
\hline
CR-B     &  CR-C  \\
\hline  						    \hline
\multicolumn{2}{|c|}{Di-tau trigger}  \\
$\ge$ 2 loose tau leptons (SS)       &   $\ge$ 2 medium tau leptons (OS)    \\
 $m(\tau_1,\tau_2) <$ 250 GeV & 	--		   	     \\
$\Delta R(\tau_1,\tau_2) >$ 1.5   &	--		              \\
$\met >$ 40 GeV 	      &  $\met >$ 40 GeV	      \\ 
20 $<\mttwo <$~50 GeV	      & 20 $<\mttwo <$~50 GeV	     \\ 
\hline
\end{tabular}

\end{minipage}
\hfillx
\begin{minipage}{0.5\textwidth}
\smallskip

\begin{tabular}{|c|c|}\hline
CR-A      & SR-D (SR-highMass)  \\
\hline
\hline
 \multicolumn{2}{|c|}{Di-tau+\met\ or asymmetric di-tau trigger} \\
$\ge$ 2 loose tau leptons (OS)       &  $\ge$ 2 medium tau leptons (OS)   \\
$<$ 1 medium tau $<$ 1 tight tau leptons	 &  $\ge$ 1 tight tau lepton	      \\
$\Delta R(\tau_1,\tau_2) >$ 1.8   & 	--		    \\
$\met >$ 110 GeV	      &  $\met >$ 110 GeV	    \\ 
$\mttwo >$ 90 GeV	      & $\mttwo >$ 90 GeV	    \\ 
\hline
\hline
VR-E    & VR-F  \\
\hline  						    \hline
\multicolumn{2}{|c|}{Di-tau or asymmetric di-tau trigger} \\
$\ge$ 2 loose tau leptons (OS)       &  $\ge$ 2 medium tau leptons (OS)   \\
$<$ 1 medium tau $<$ 1 tight tau leptons	 &  $\ge$ 1 tight tau lepton	      \\
$\Delta R(\tau_1,\tau_2) >$ 1.8   &	--			\\
$\met >$ 40 GeV 	      &  $\met >$ 40 GeV	    \\ 
60 $<\mttwo <$~90 GeV	      & 60 $<\mttwo <$~90 GeV	    \\ 
\hline
\hline
CR-B     &  CR-C  \\
\hline  						    \hline
\multicolumn{2}{|c|}{Di-tau or asymmetric di-tau trigger} \\
$\ge$ 2 loose tau leptons (OS)       &  $\ge$ 2 medium tau leptons (OS)   \\
$<$ 1 medium tau $<$ 1 tight tau leptons	 &  $\ge$ 1 tight tau	       \\
$\Delta R(\tau_1,\tau_2) >$ 1.8   &	--			\\
$\met >$ 40 GeV 	      &  $\met >$ 40 GeV	    \\ 
10 $<\mttwo <$~60 GeV	      & 10 $<\mttwo <$~60 GeV	    \\ 
\hline
\end{tabular}
\end{minipage}
\label{tab:QCD-CR}
\end{table}

\begin{figure}[htpb]
\centering
\includegraphics[width=0.495\textwidth]{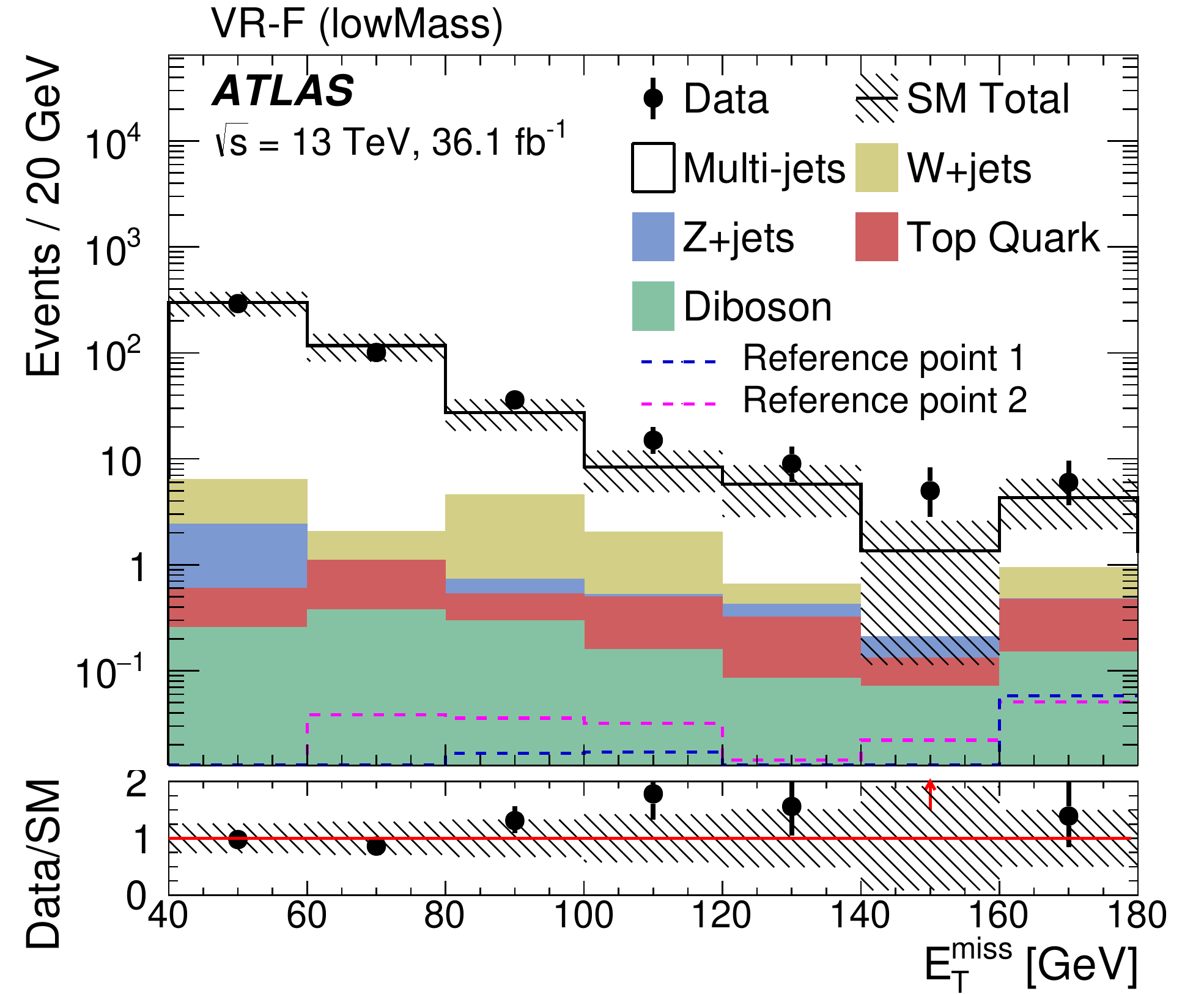}
\includegraphics[width=0.495\textwidth]{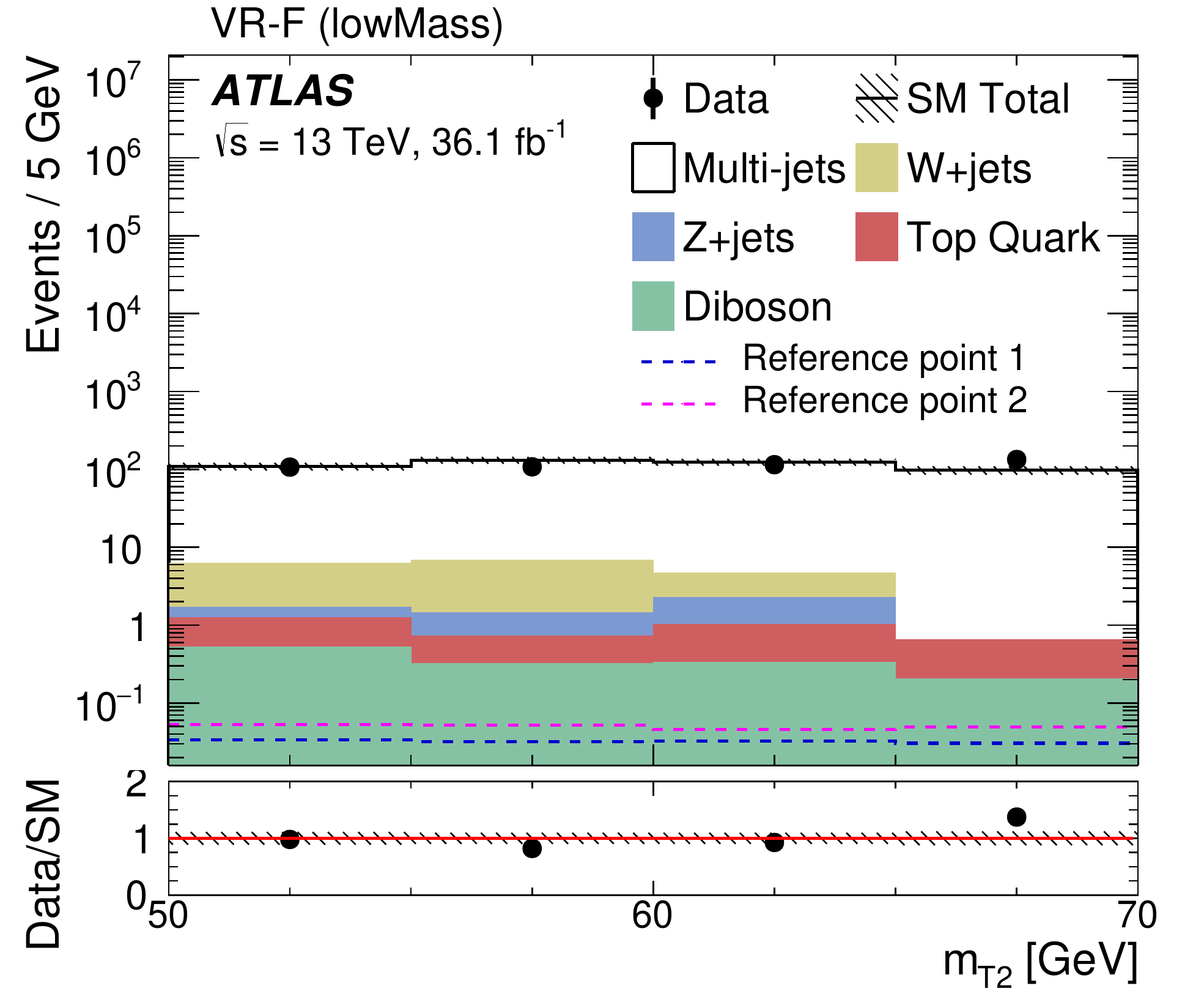}\\
\includegraphics[width=0.495\textwidth]{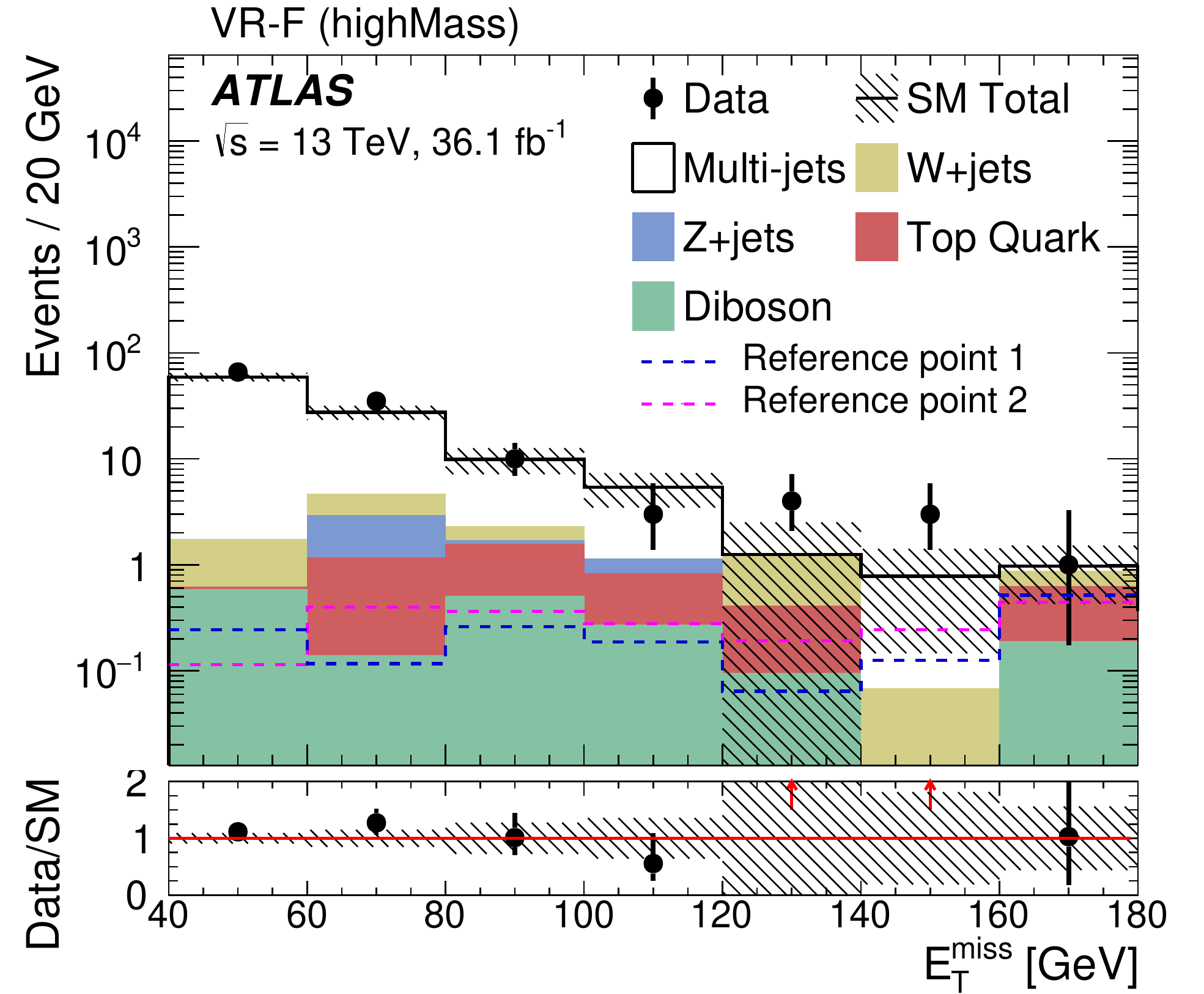}
\includegraphics[width=0.495\textwidth]{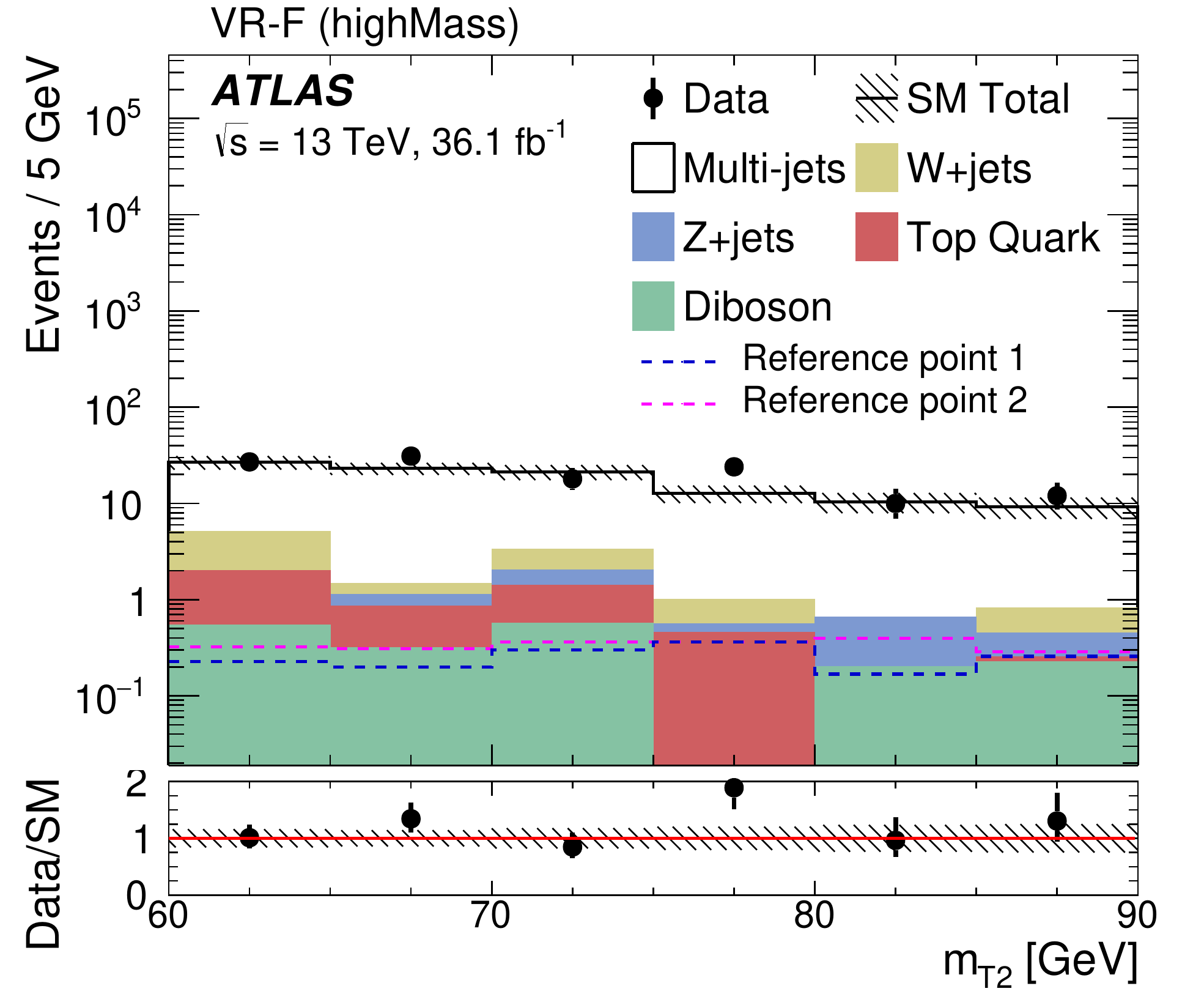}\\

\caption{The \met\ (left) and \mttwo\ (right) distributions in the
  multi-jet  background VR-F for SR-lowMass (top) and VR-F for SR-highMass (bottom). 
The stacked histograms show the contribution of the non-multi-jet SM backgrounds from 
MC simulation.
The multi-jet contribution is
estimated from data using the ABCD method. The hatched bands represent
the combined statistical and systematic uncertainties in the sum of the SM backgrounds shown. 
For illustration, the distributions of the SUSY reference points (defined in Section \ref{sec:data}) 
are also shown as dashed lines. The last bin in the left panels includes the overflow events.
}
\label{fig:MET-regionAB}
\end{figure}

The number of multi-jet events in the control and validation regions is estimated from data after
subtraction of other SM contributions estimated from MC simulation.
In both CR-B and VR-E,  more than \percent{86} of the events come from multi-jet production, whereas for   
CR-A and CR-C  the multi-jet purity is larger than \percent{47} and \percent{68}, respectively. In VR-F the multi-jet purity is larger than \percent{90}.
Agreement between data and the estimated SM background is found for the \met\
and \mttwo\ distributions in the validation regions, as shown in Figure \ref{fig:MET-regionAB}.
The correlation between the tau identification and charge and the kinematic variables is checked by studying the variation of the 
transfer factor $\mathrm{T}$ as a function of the kinematic variables \mttwo\ and \met, and is found to be negligible. 
The results of the ABCD method are summarised in Table~\ref{tab:qcd-est-result}. 

The signal contamination in a certain region is defined as the ratio of the number of signal events to the sum of the number of signal events and SM background processes. 
The signal contamination in CR-A for both SRs ranges from a few percent to \percent{30--50} for a few signal models, and 
it is taken into account in the simultaneous fit described in Section \ref{sec:fit}. 
The largest contaminations are found for a \chinoonepm\ mass of 400~GeV and massless \ninoone\ for \ConepConem production,
and for a \chinoonepm\ mass of 300~GeV and massless \ninoone\ for \ConepmNtwo production. 
The possible presence of non-SM event contamination in CR-A was tested and
proved not to change the fit results significantly.

\begin{table}
\centering
\caption{  
The MC predicted backgrounds in the multi-jet control regions, including the statistical uncertainties,  and the expected multi-jet contribution (in italics), 
obtained by subtracting the MC contributions from observed data (in bold).  
Predicted event yields for the SUSY reference points (defined in Section~\ref{sec:data}) in the control regions are also shown.
The estimated multi-jet contribution in the SRs is given in the last column including both the statistical and systematic uncertainties. 
The details of the systematic uncertainties reported here are discussed in Section \ref{sec:Systematics}. 
}
\small
\vspace{0.5 cm}
\begin{tabular}{|c|l|c|c|c|c|c|}\hline
  SR	      &  Sample 	     & CR-B       & CR-C    &  CR-A   & $\mathrm{T}$ = C/B	 & Multi-jet  in SR-D\\ 
\hline\hline  
	     &$\mathbf {Data }$   &	 $\mathbf { 556 }$	 & $\mathbf { 674 }$	   &  $\mathbf { 8 } $        &        &  \\ \cline{2-5}   
	     &$Z$+jets    & 3.4 $\pm$ 2.1 & 19 $\pm$ 5\hd & 0.8 $\pm$ 0.4       &	  &	   \\ \cline{2-5} 
	     &$W$+jets    & 8.9 $\pm$ 1.8 & 20 $\pm$ 5\hd & 1.8 $\pm$ 1.0      & 	    &	     \\ \cline{2-5}
lowMass        &Diboson     & 0.94 $\pm$ 0.12 & 3.3 $\pm$ 0.2 & 0.29 $\pm$ 0.07	 & 1.16   & 4.3   \\ \cline{2-5}
	     &Top	    & 1.61 $\pm$ 0.30 & 4.7 $\pm$ 0.5 & 1.4 $\pm$ 1.1		  & $\pm$ 0.07& $\pm$ 4.0 \\ \cline{2-5}    
&\textit {Multi-jet}  	&\textit{541 $\pm$ 24}\hd& \textit{627 $\pm$ 27}\hd  & \textit {3.7 $\pm$ 1.6}     &	&      \\ \cline{2-5}
	     &Reference point 1 & 0.06 $\pm$ 0.01 & 0.16 $\pm$ 0.02 & 1.68 $\pm$ 0.16      &	      &        \\ \hline \hline 	    
	     &$\mathbf {Data }$   &	 $\mathbf { 1565 }$	  & $\mathbf { 836 }$	    &  $\mathbf { 5 } $        &	&  \\ \cline{2-5}	
	     &$Z$+jets	    & 56 $\pm$ 31 & 93 $\pm$ 42 & 0.02 $\pm$ 0.29   &		  &	   \\ \cline{2-5}      
	    &$W$+jets 	   & 151 $\pm$ 22\hd & 125 $\pm$ 17\hd & 1.1 $\pm$ 0.4       &	   &	    \\ \cline{2-5} 
highMass	 &Diboson		& 9.6 $\pm$ 1.1 & 20.5 $\pm$ 2.0\hd & 0.8 $\pm$ 0.4      &  0.43 & 1.3   \\ \cline{2-5}
	     &Top		    & 9.2 $\pm$ 1.5 & 25.4 $\pm$ 3.4\hd & 0.01 $\pm$ 0.01     & $\pm$ 0.04 & $\pm$ 1.1 \\ \cline{2-5}
             &\textit{Multi-jet}  	&\textit{1340 $\pm$ 50}\hd\hd& \textit { 570 $\pm$ 50 }\hd  & \textit{3.1 $\pm$ 0.6}    &        &      \\ \cline{2-5}
	     &Reference point 2	    & 0.53 $\pm$ 0.08 & 2.37 $\pm$ 0.21 & 1.92 $\pm$ 0.16	 &	      &        \\ \hline 
\end{tabular}
\label{tab:qcd-est-result}
\end{table}

\subsection{$W$+jets background estimation}
\label{sec:BG_W}

The production of $W$+jets events with at least one misidentified tau lepton
is an important background, accounting for about \percent{13} (\percent{20}) of the expected SM background in
SR-lowMass (SR-highMass).
A dedicated control region ($W$-CR) is used to normalise the $W$+jets MC estimate to data.
To suppress multi-jet contamination, the $W$-CR is enriched in events
where the $W$ boson decays leptonically into a muon and a neutrino. Events are selected with a single-muon trigger, using the lowest unprescaled $\pT$ thresholds available.
Events containing exactly one isolated muon and one candidate tau lepton with opposite electric charge are selected.
The muon is required to have \pt~$>$ 40 GeV.
In addition, the muon must satisfy the `GradientLoose'~\cite{PERF-2015-10} isolation requirements, which rely on the use of track and 
calorimeter based variables and implement a set of $\eta$- and $\pt$-dependent criteria. 
Compatibility of the signal lepton tracks with the primary vertex is enforced by
requiring $|z_0\sin\theta| < 0.5~\mathrm{mm}$, where $z_0$ is the longitudinal impact
parameter. In addition, the transverse impact parameter, $d_0$,
divided by its uncertainty, $\sigma(d_0)$, must satisfy
$|d_0/\sigma(d_0)|<3$ for the muon.
The tau candidate must satisfy the medium tau identification criteria and is required to have \pt $>$ 50 GeV.

The contribution from events with top quarks is suppressed by rejecting events containing $b$-tagged jets.
To reduce the contribution from $Z$+jets production, the transverse
mass of the $\mu+\met$ system, $m_{\mathrm{T},\mu} >$~50~GeV,  the
sum of the transverse mass of the $\tau+\met$ and $\mu+\met$ systems, $m_{{\mathrm T},\tau}
+ m_{{\mathrm T},\mu} >$~80~GeV, and the angular separation between the muon
and the tau lepton $\Delta R(\mu,\tau)>$ 0.5 are required. To further suppress diboson
and top-quark contributions, $m_{\mathrm{T},\mu} <$~150~GeV is required.
To be close to the SR definition, $\met >$ 60 GeV and the invariant mass of the muon and tau lepton, $m(\mu, \tau) >$ 70 GeV are required.
Events in the $W$-CR are selected by requiring low \mttwo, while a high \mttwo\ region is used to validate 
the $W$+jets estimate ($W$ validation region, $W$-VR).
The definitions of the $W$-CR and $W$-VR are given in  Table \ref{tab:WCR}.
 
The multi-jet contribution in the $W$-CR ($W$-VR) is estimated using the
so-called \textit{OS--SS} method by counting the number of events in data 
satisfying the same requirements as the $W$-CR ($W$-VR) but with the
electric charge of the two leptons having the same sign (SS). 
Events from SM processes other than multi-jet production are
subtracted from the data counts in the SS region using MC simulation.
The OS--SS method relies on the fact that in the multi-jet background the ratio of SS to OS events is close to unity, while a significant difference from 
unity is expected for $W$+jets production. The latter is dominated by $gu/gd$-initiated processes that often give rise to a jet originating from a quark, the 
charge of which is anti-correlated with the $W$ boson charge.
Based on studies with simulated samples, a systematic
uncertainty of \percent{100} is assigned to the multi-jet estimate in
the $W$-CR.

\begin{figure}
  \centering
\includegraphics[width=0.495\textwidth]{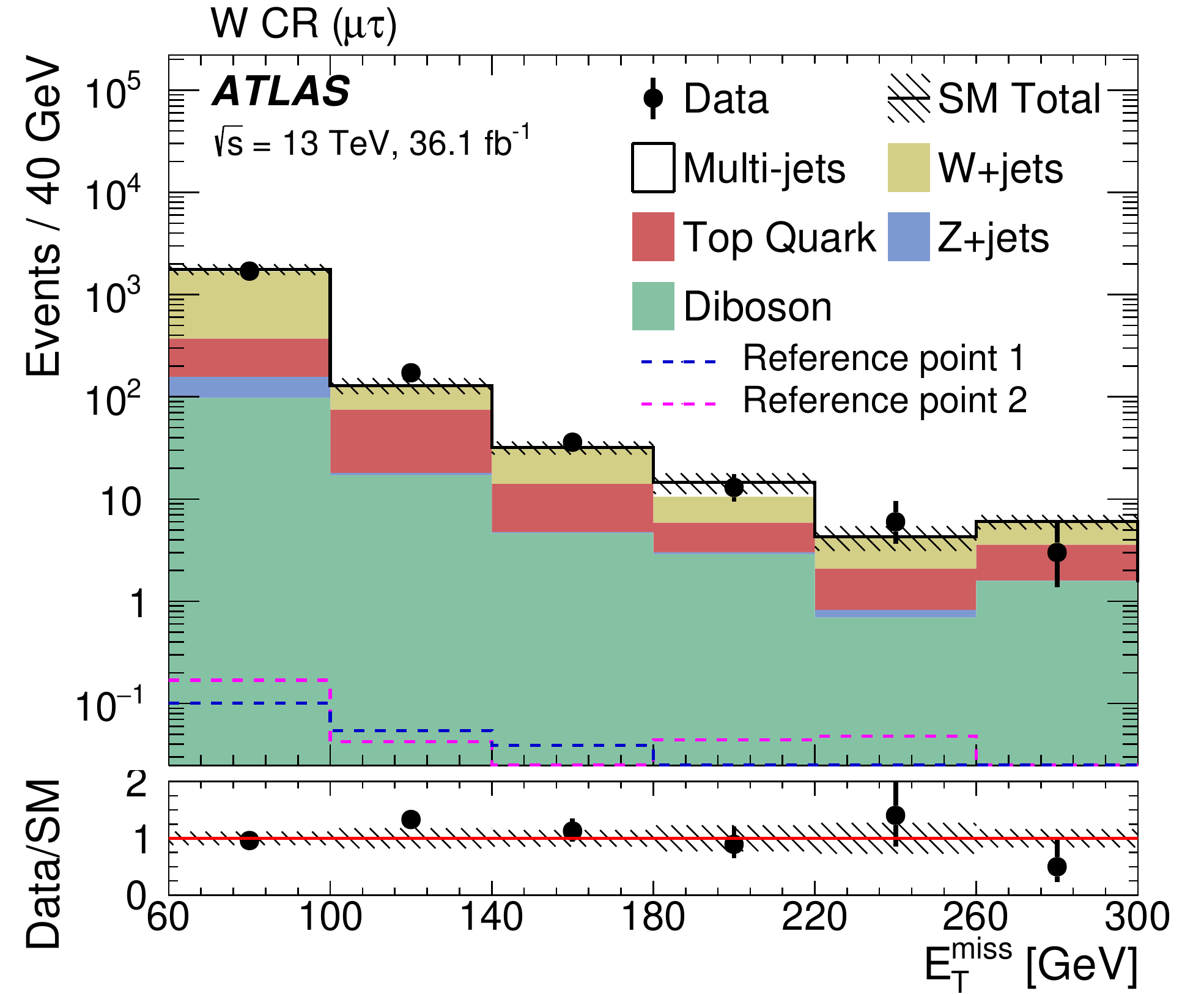}
\includegraphics[width=0.495\textwidth]{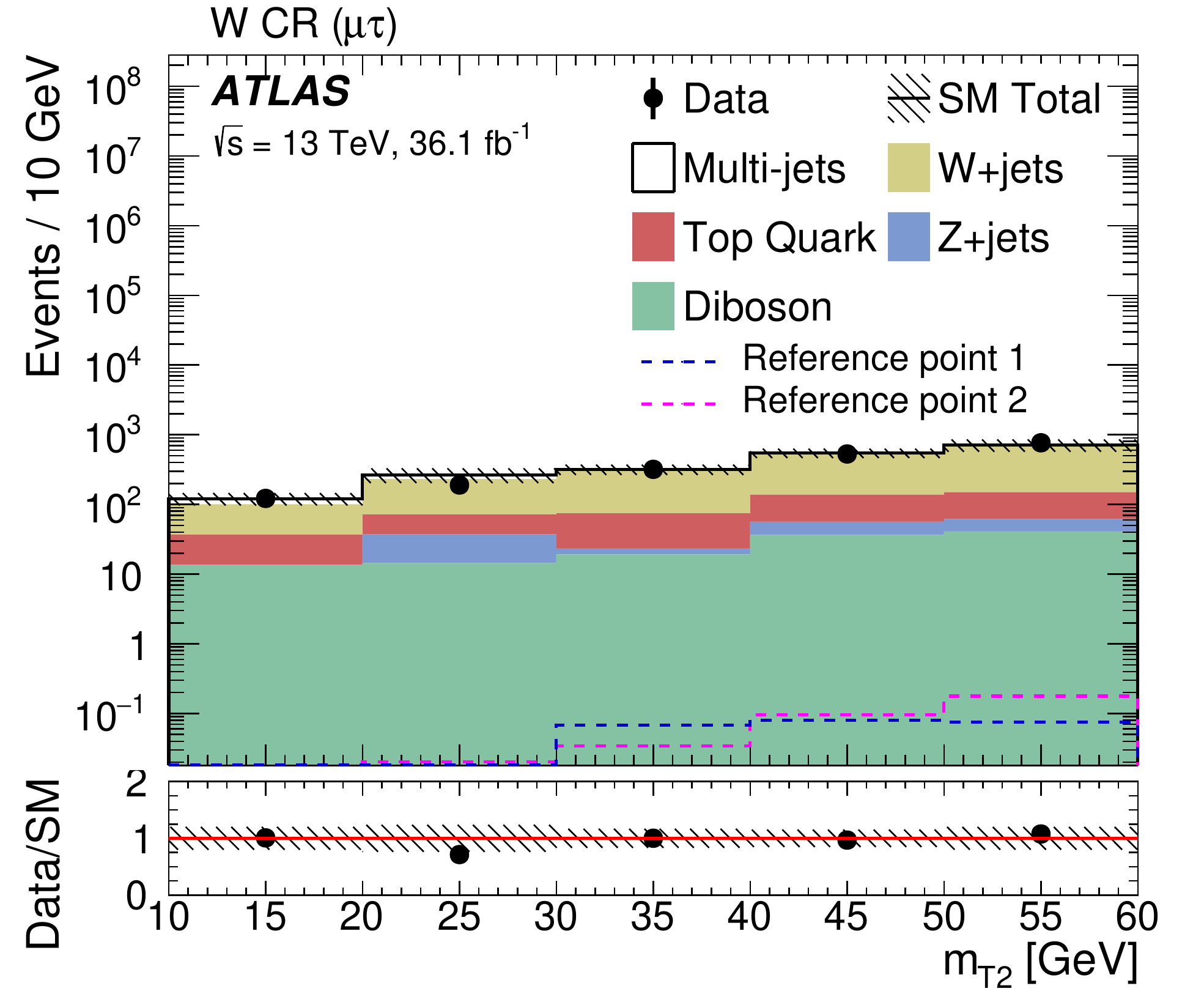}\\
\includegraphics[width=0.495\textwidth]{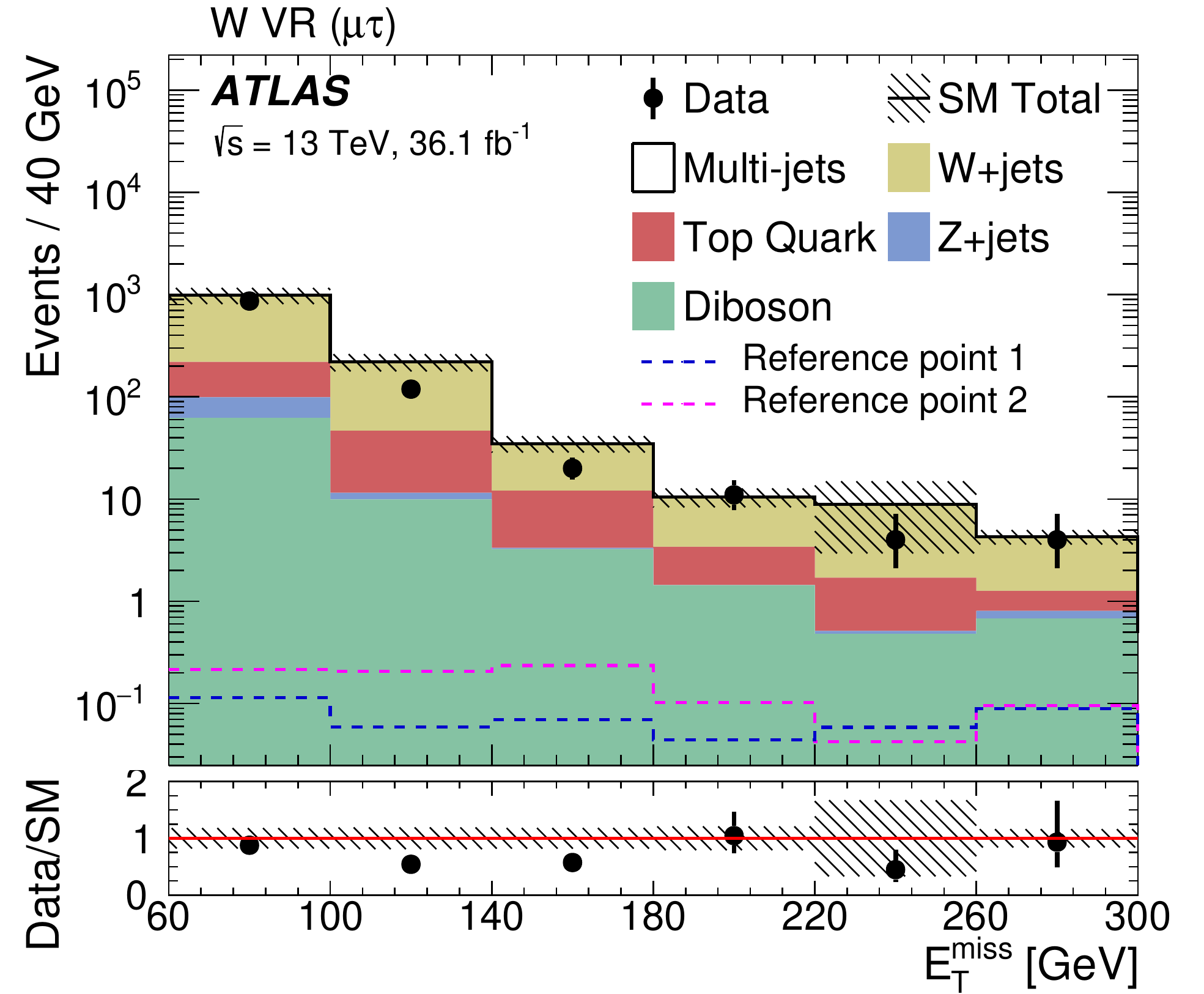}
\includegraphics[width=0.495\textwidth]{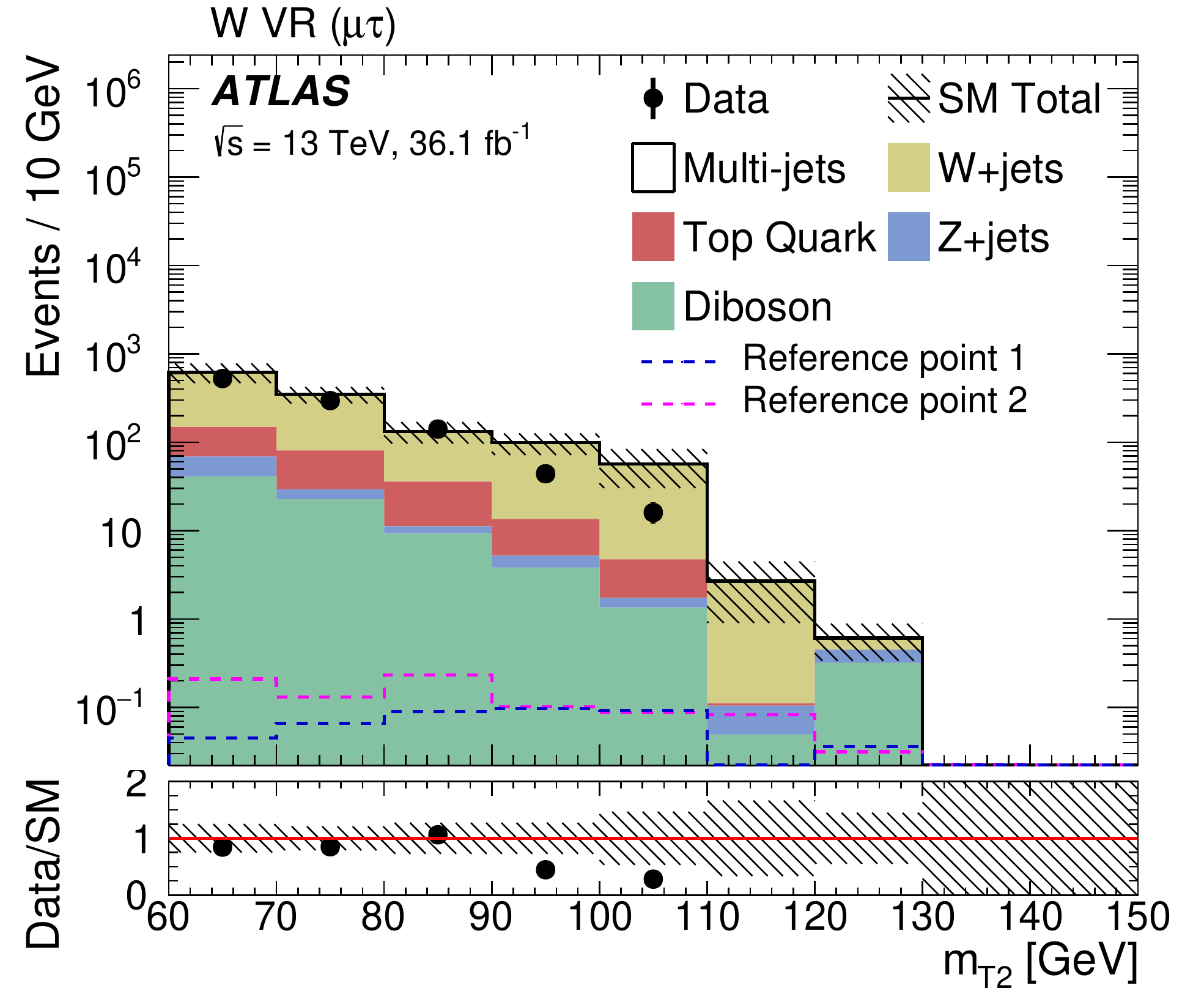}\\
  \caption{
    The \met\ (left) and \mttwo\ (right) distributions in the $W$-CR
    (top) and $W$-VR (bottom) regions.    
    The SM backgrounds other than multi-jet production are estimated from MC simulation.  
    The contribution of $W$+jets events is scaled to the fit result. 
    The multi-jet contribution is estimated from data using the \textit{OS--SS} method.  
    The hatched bands represent the combined statistical and systematic uncertainties of the total SM background. 
    For illustration, the distributions of the SUSY reference points defined in Section \ref{sec:data} are also shown as dashed lines. 
    The lower panels show the ratio of data to the SM background
    estimate. The last bin includes the overflow events.
  }
  \label{fig:WCR}
\end{figure}

The event yields in the $W$-CR and $W$-VR are given in Table \ref{tab:WCRVR}. 
The purity of the selection in $W$+jets events is around \percent{72} (\percent{77}) in the $W$-CR ($W$-VR). Agreement between data and SM predictions is
observed. The signal contamination in the $W$-CR and $W$-VR is negligible.
Distributions of the kinematic variables defining the SRs are shown in
Figure \ref{fig:WCR}, in which the contribution of $W$+jets events is scaled
with the normalisation factor 1.02 obtained from the fit described in Section \ref{sec:fit}. 
The discrepancy between observed data and predictions at $\mttwo\ >$ 90 GeV in the $W$-VR is due to events with different kinematics from the SRs, with either $\met <$ 150 GeV or where the muon has \pt~$<$ 60 GeV.

\begin{table}[!hbp]
\caption{\label{tab:WCR} The $W$-CR (left) and $W$-VR (right) definitions.}
\begin{center}
\begin{tabular}{|c|c|}
\hline 
 $\hd\hd\hd\hd\hd\hd\hd\hd\hd\hd$  $W$-CR $\hd\hd\hd\hd\hd\hd\hd\hd\hd$                & $W$-VR \\
\hline                              
\hline                              
\multicolumn{2}{|c|}{One isolated muon and one medium tau lepton with opposite sign}   \\                                                                      
\multicolumn{2}{|c|}{$b$-jet veto}                                   \\
\multicolumn{2}{|c|}{$m(\mu, \tau) >$ 70 GeV }\\
\multicolumn{2}{|c|}{$\met >$ 60 GeV}                                \\
\multicolumn{2}{|c|}{ 50 $< m_{\mathrm{T},\mu} <$~150~GeV }                   \\
\multicolumn{2}{|c|}{ $m_{\mathrm{T},\mu}+m_{\mathrm{T},\tau} >$ 80 GeV}               \\
0.5 $< \Delta R(\mu,\tau) <$ 3.5  & 0.5 $< \Delta R(\mu,\tau) <$ 4.5 \\
 10 $< \mttwo\ <$ 60 GeV           & $\mttwo\ >$ 60 GeV               \\
\hline
\end{tabular}
\end{center}
\end{table}

\begin{table}[!h]
 \caption{ 
  Event yields in the $W$-CR and $W$-VR.
  The SM backgrounds other than multi-jet production are estimated from MC simulation. 
  The contribution of $W$+jets events is scaled with the normalisation
  factor obtained from the fit. 
  The multi-jet contribution is estimated from data using the OS--SS
  method. In the $W$-VR the multi-jet estimation with the OS-SS method
  yields a negative contribution, which is set to zero.
Predicted event yields for the SUSY reference points (defined in Section~\ref{sec:data}) are also shown.
The uncertainties given are the sum in quadrature of statistical and systematic uncertainties. 
The correlation of systematic uncertainties among control and validation regions and background processes
is fully taken into account in the fit.
  }
  \smallskip
 \centering
 \begin{tabular}{|l|c|c|}
 
   \hline
   Sample   & $W$-CR                      & $W$-VR      \\
\hline
\hline
  Data	       & $1928$             	& $1023$   \\
 SM total      & $1930 \pm 50$\hd\hd  & $1260 \pm 440$\hd \\
\hline
   $W+$jets    & $1395 \pm 130$\hd & $980 \pm 410$  \\
\hline
   $Z+$jets    & $60 \pm 28$    & $39 \pm 15$    \\
   Diboson     & $125 \pm 24$\hd   & $78 \pm 20$    \\
   Top quark   & $290 \pm 80$\hd   & $170 \pm 60$\hd   \\
   Multi-jet   & $60 \pm 60$        &  \hd\hd  $0 \pm 100$    \\
\hline
 Reference point 1   & $  0.22 \pm  0.07$  & $ 0.44 \pm  0.08$ \\
 Reference point 2   & $  0.33 \pm  0.08$  & $  0.87 \pm  0.11$\\
\hline
\end{tabular}
\label{tab:WCRVR}
\end{table}

\subsection{Irreducible background estimation}
\label{sec:BG_irre}

Irreducible SM backgrounds arise mainly from \ensuremath{t\bar{t}},
single top quark, \ensuremath{t\bar{t}}+$V$, $Z$+jets, and diboson ($WW$, $WZ$ and $ZZ$) processes and
are estimated with MC simulation. Other SM backgrounds are negligible. 

The inclusive contribution from \ensuremath{t\bar{t}}, single top quark, \ensuremath{t\bar{t}}+$V$ and $Z$+jets
amounts to about \percent{18} (\percent{13}) of the total background in SR-highMass (SR-lowMass).
The MC estimates are validated in regions enriched in $Z$+jets and top-quark events. 
For both regions, the events passed either the combined
di-tau + \met trigger or the asymmetric di-tau trigger. Events are required to have 
at least two tau candidates with opposite electric charge, 
\met~$>$~150 GeV, and leading (sub-leading) tau $p_{\mathrm T} >$ 50 (40) GeV.
In the $Z$+jets validation region ($Z$-VR), at least two tau
candidates must satisfy the medium tau identification criteria. 
To suppress top-quark backgrounds, events containing $b$-tagged jets are vetoed.
To further enhance the purity of $Z$+jets events, \mttwo~$<$~10 GeV is required.
In the top-quark validation region (Top-VR), at least one tau candidate
must satisfy the medium tau identification criteria. 
To increase the contribution from top-quark events, events must contain at least one $b$-tagged jet with $p_{\mathrm T} > 20$ GeV and must be
kinematically compatible with \ttbar\ production (top-tagged) 
through the use of the \textit{contransverse} mass $m_{\mathrm {CT}}$ \cite{Tovey:2008}.
The scalar sum of the $p_{\mathrm T}$ of the two tau leptons and of at least one
combination of two jets in an event must exceed 100 GeV. Top-tagged events are required to possess $m_{\mathrm {CT}}$
values calculated from combinations of jets and tau leptons consistent with the expected bounds from $t\bar{t}$ events as described in
Ref.~\cite{Polesello:2009rn}. 
The $Z$-VR and Top-VR requirements are summarised in Table \ref{tab:ZTvr}.

The diboson background accounts for \percent{26} (\percent{43}) of the
total SM contribution in the SR-highMass (SR-lowMass) and mainly arises
from $WW\rightarrow \tau\nu \tau\nu$  and $ZZ \rightarrow \tau\tau
\nu\nu$ events, in which more than \percent{96} of the contribution is
from events with two real tau leptons according to the MC simulation.
To validate the MC modelling and normalisation of the $WW$ ($ZZ$)
process, a validation region $WW$-VR ($ZZ$-VR) with an enriched $WW
\rightarrow e \nu \mu\nu$ ($ZZ \rightarrow e e \nu \nu$ or $ZZ \rightarrow \mu\mu \nu \nu$)
contribution is defined.
For $WW$-VR, events with two isolated leptons ($\ell = e$ or $\mu$) with different flavour and
opposite sign are selected, while for $ZZ$-VR, events with two isolated 
leptons with same flavour and opposite sign are selected.
To keep the phase space similar to the SRs, $WW$-VR ($ZZ$-VR) is defined
to be close to the SRs except for the selected objects
being a light-lepton pair. Top-tagged events are vetoed to suppress
the \ttbar contribution in $WW$-VR. To suppress the $Z$+jets contribution in
$ZZ$-VR,  $\Delta R(\ell,\ell) <$ 1.5 is applied; the requirement 
$|m_{\ell\ell}-m_Z| <$~15 GeV is used to enrich the $ZZ$ contribution.
The definitions of $WW$-VR and $ZZ$-VR are summarised in Table \ref{tab:VVvr}.

\begin{figure}[htpb]
\centering
\includegraphics[width=0.495\textwidth]{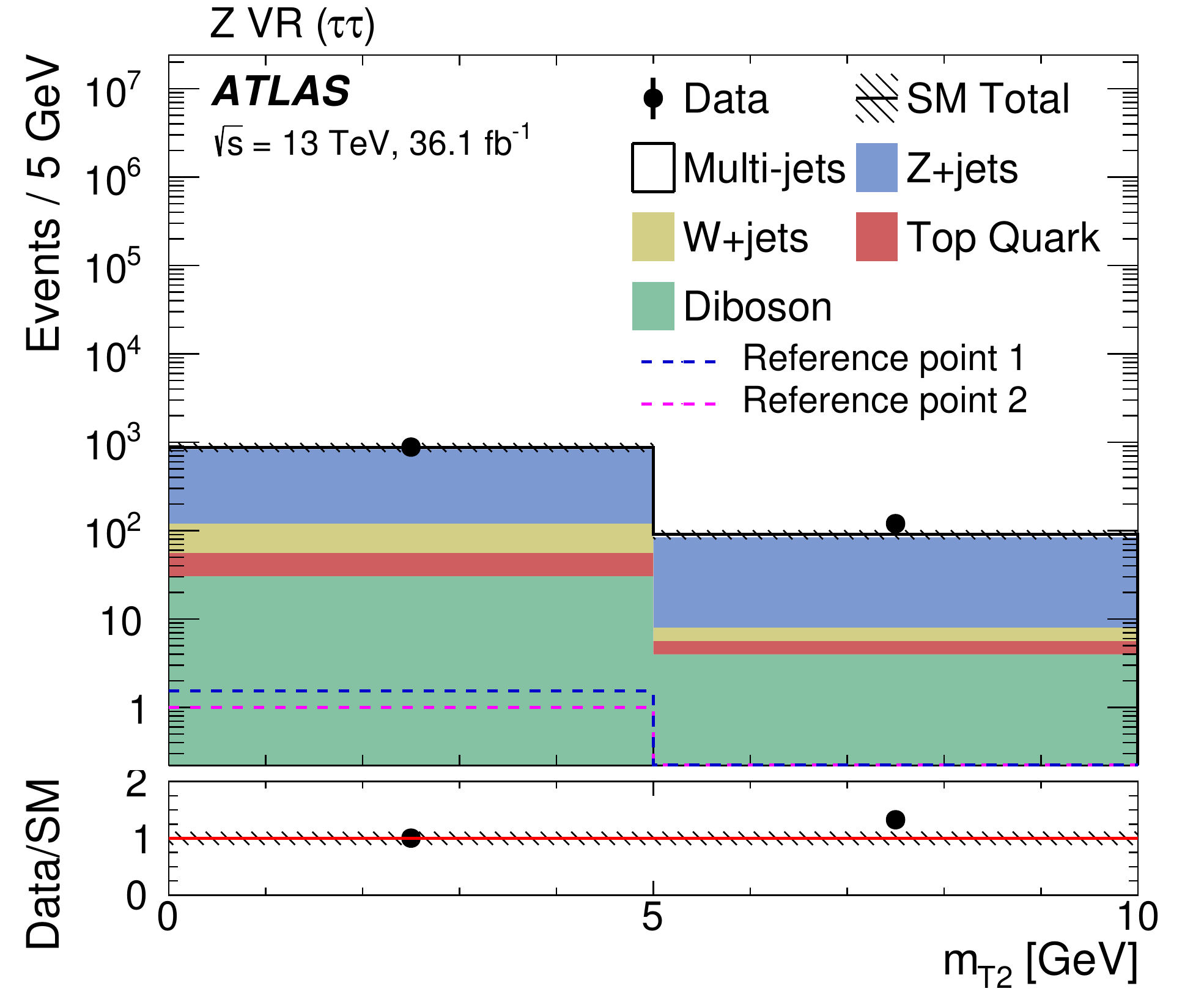}
\includegraphics[width=0.495\textwidth]{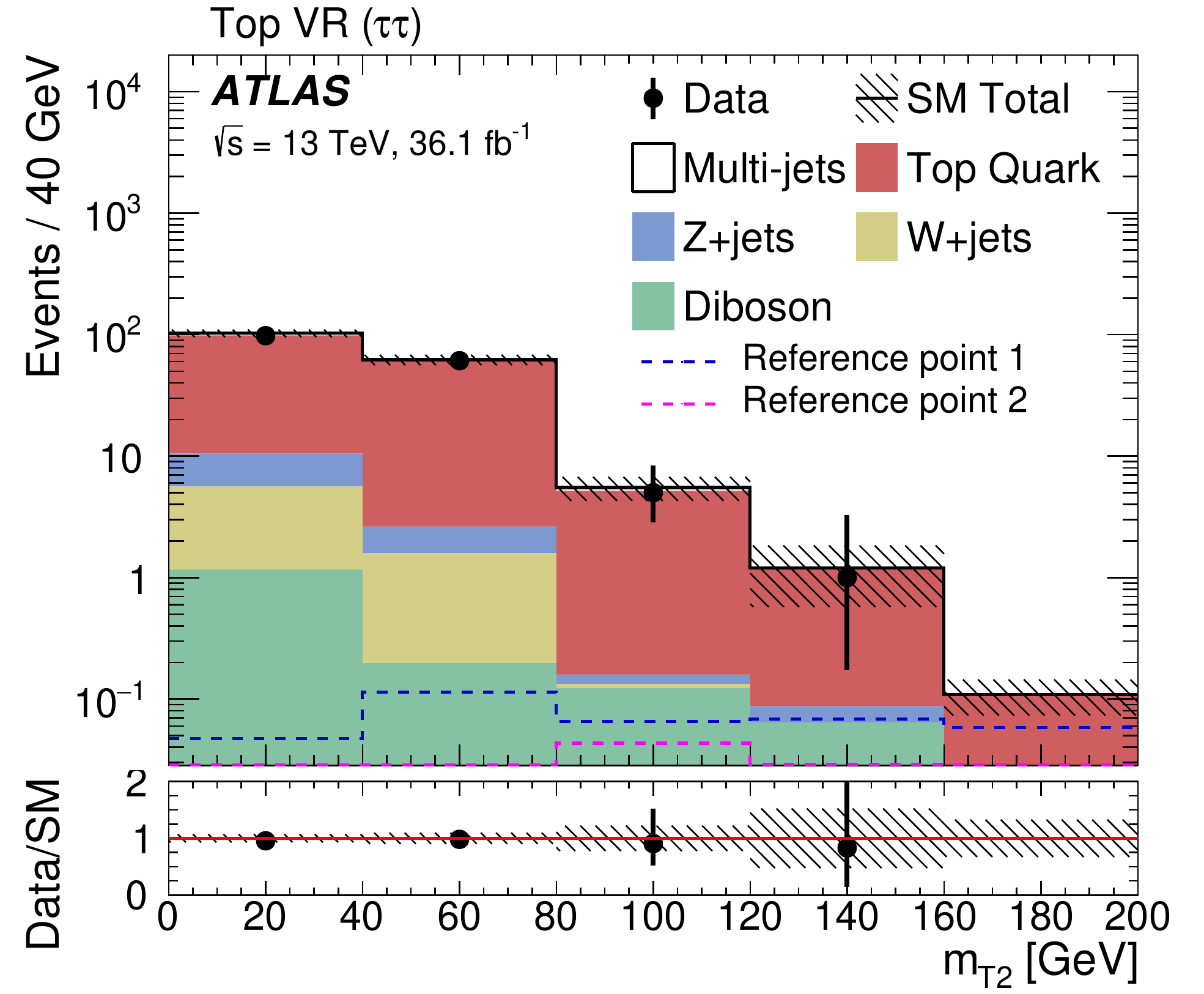}\\
\includegraphics[width=0.495\textwidth]{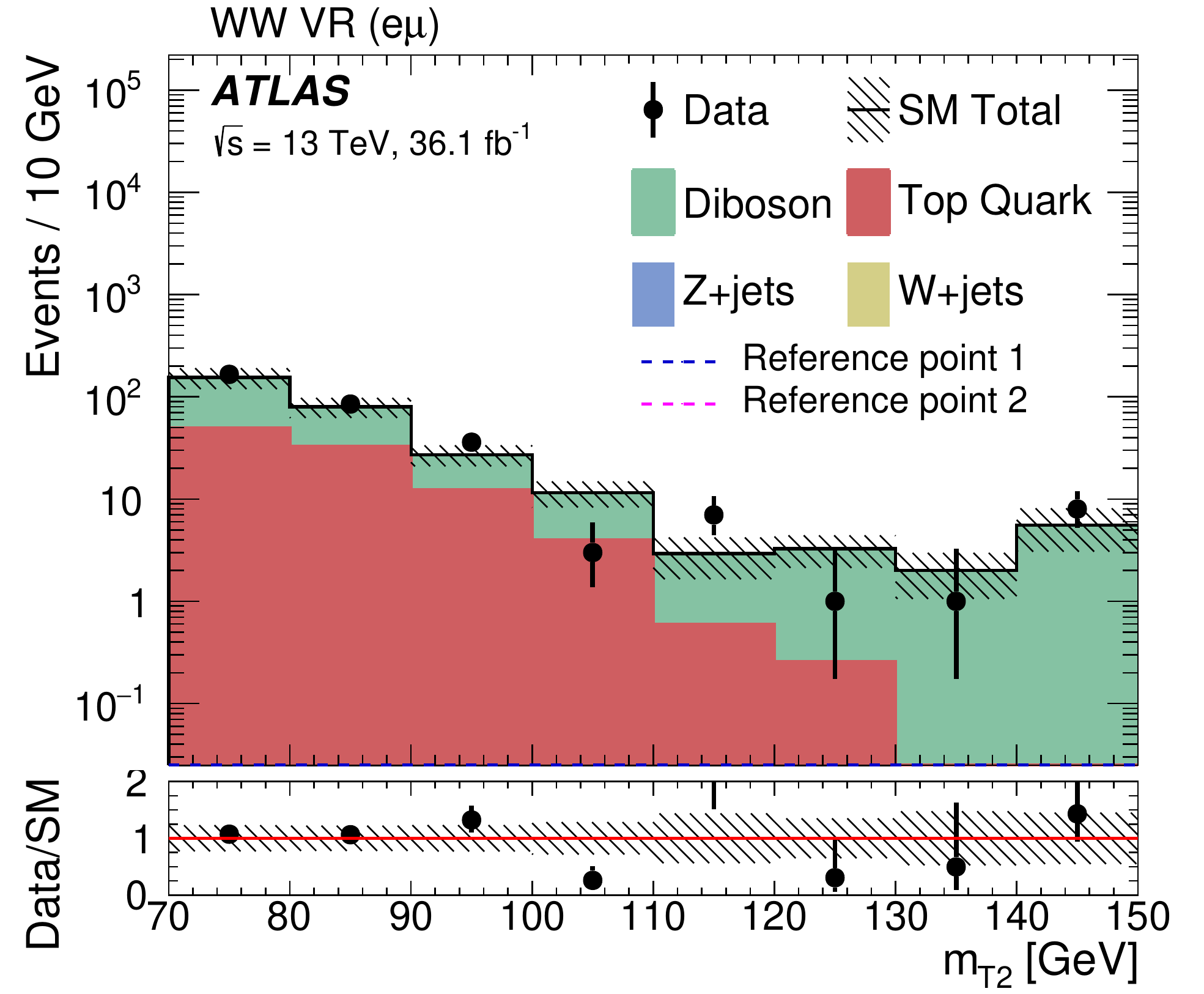}
\includegraphics[width=0.495\textwidth]{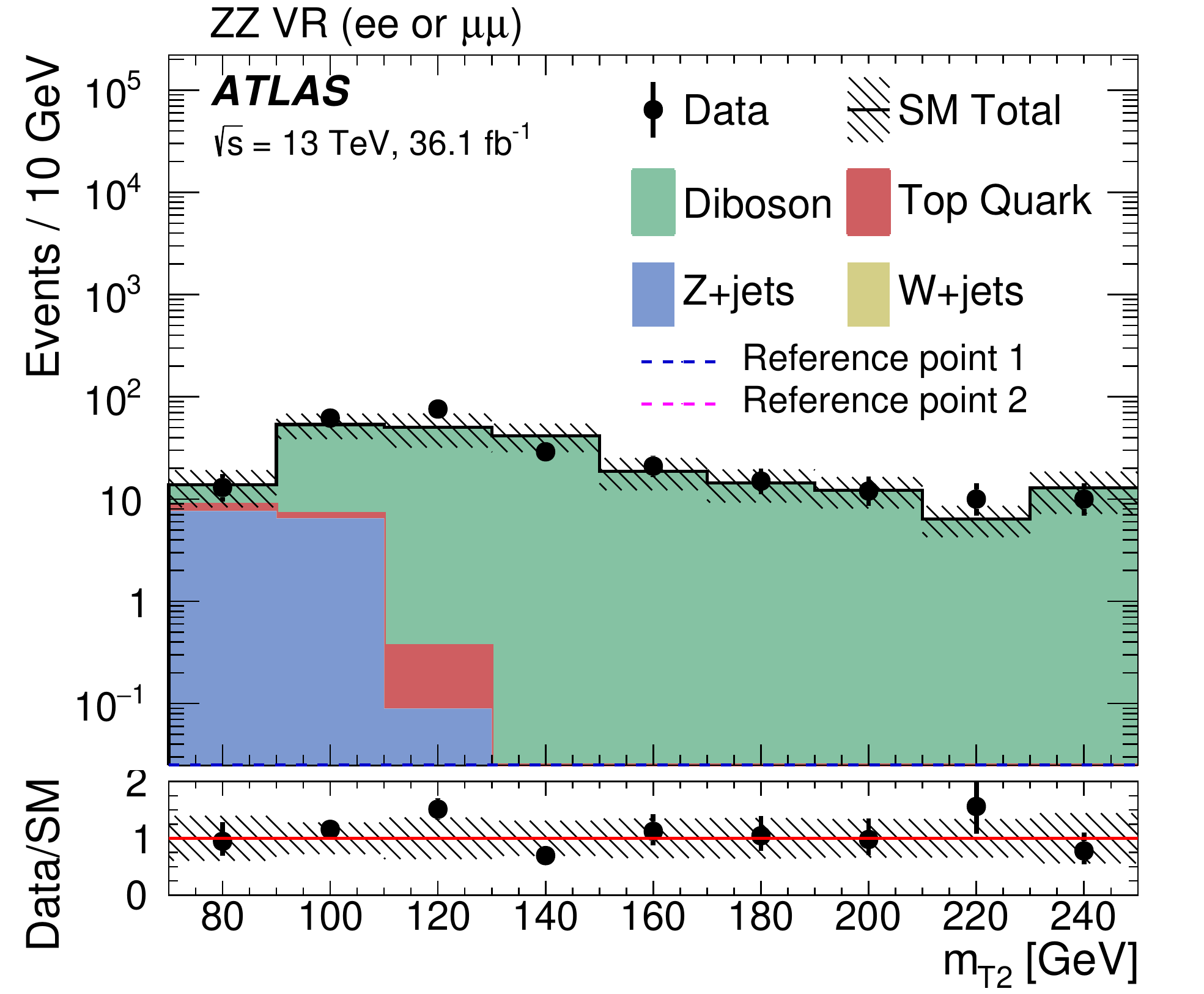}\\

\caption{The \mttwo\ distribution in the $Z$-VR (top left), Top-VR
  (top right), $WW$-VR (bottom left) and $ZZ$-VR (bottom right) regions.
The SM backgrounds other than multi-jet production are estimated from MC simulation. 
The multi-jet contribution is
negligible and not considered in $WW$-VR and $ZZ$-VR, while in $Z$-VR and Top-VR it is
estimated from data using the ABCD method, using CRs obtained with the
same technique used for the SRs, and described in Section \ref{sec:BG_QCD}. The hatched bands represent
the combined statistical and systematic uncertainties of the total SM background. 
For illustration, the distributions of the SUSY reference points (defined in Section \ref{sec:data}) are
also shown as dashed lines. The lower panels show the ratio of data to
the SM background estimate.  The last bin includes the overflow events
except for the upper left panel.
}
\label{fig:bkg-vr}
\end{figure}

The purity of the selection in $Z$+jets and \ttbar\ events is above \percent{80} in the respective
validation regions, and the purity of the selection in $WW$ ($ZZ$) events is around \percent{65} (\percent{92}) in $WW$-VR ($ZZ$-VR).
Agreement between data and the SM prediction is observed in all validation regions.
The \mttwo\ distributions in the $Z$-VR, Top-VR, $WW$-VR and $ZZ$-VR are shown in Figure \ref{fig:bkg-vr}.

\begin{table}[!hbp]
\caption{\label{tab:ZTvr} The $Z$-VR (left) and Top-VR (right) definitions.}
\begin{center}
\begin{tabular}{|c|c|}
\hline 
 $Z$-VR                 & Top-VR \\
\hline                              
\hline                              
\multicolumn{2}{|c|}{At least one opposite-sign tau lepton pair}\\
\multicolumn{2}{|c|}{Tau \pt$>$ 50, 40 GeV} \\
\multicolumn{2}{|c|}{$\met >$ 60 GeV} \\
At least two medium tau leptons      &  at least one medium and one loose tau lepton \\
$b$-jet veto       & at least one $b$-jet \\
$\mttwo\ <$ 10 GeV &  $\mttwo\ >$ 10 GeV            \\
--                 &  $m_{\mathrm {CT}}$ top-tagged \\
\hline
\end{tabular}
\end{center}
\end{table}

\begin{table}[!h]
\caption{\label{tab:VVvr} The $WW$-VR (left) and $ZZ$-VR (right) definitions.}
\begin{center}
\begin{tabular}{|c|c|}
\hline
 $WW$-VR                 & $ZZ$-VR \\
\hline
\hline
\multicolumn{2}{|c|}{One opposite-sign lepton ($e$ or $\mu$) pair}   \\
\multicolumn{2}{|c|}{$\mu$ \pt~$>$30 GeV, $e$ \pt $>$ 40 GeV} \\
\multicolumn{2}{|c|}{Jet veto}\\
\multicolumn{2}{|c|}{$m_{\ell\ell}~>$~50 GeV } \\
\multicolumn{2}{|c|}{$\met >$ 50 GeV} \\
\multicolumn{2}{|c|}{ $m_{\mathrm{T}, \mu} >$ 100 GeV} \\
\multicolumn{2}{|c|}{  $\mttwo\ >$ 70 GeV} \\
Two isolated leptons ($e$ or $\mu$) with different flavour & two isolated leptons ($e$ or $\mu$) with same flavour  \\
 $m_{\mathrm {CT}}$ top tag veto & $\Delta R(\ell,\ell) <$ 1.5 \\
--   &  $|m_{\ell\ell}-m_Z|$~$<$~15 GeV \\\hline
\end{tabular}
\end{center}
\end{table}

\subsection{Statistical analysis}
\label{sec:fit}

The statistical interpretation of the results is performed using the profile likelihood method implemented in the 
HistFitter framework~\cite{Baak:2014wma}. Three types of fits are performed for each SR.

\begin{itemize}
\item
The \textit{background-only} fit uses as input the number of observed events in the multi-jet CR-A and $W$-CR, the expected SM contributions other than
multi-jet to the multi-jet CR-A and $W$-CR, and the transfer factors,
which are used to extrapolate the background of multi-jet or $W$+jets events in their 
control regions to these predicted in the signal regions.
The free parameters in the fit are the normalisations of the $W$+jets and multi-jet contributions.
The signal is assumed to be absent in this fit.
\item
A \textit{model-independent limit} fit combines the data event yield in a given SR with the SM background estimate
and its uncertainties obtained by the background-only fit to test whether any non-SM signal contributes to the SR.
The significance of a possible excess of observed events over the SM prediction 
is quantified by the one-sided probability, $p (\text{signal} = 0)$ denoted by $p_0$, of the background alone to fluctuate to the observed number of events or higher
using the asymptotic formula described in Ref.~\cite{statforumlimitsl}. The
presence of a non-SM signal would manifest itself in a small $p_0$ value. 
\item In the \textit{model-dependent limit} fit the SUSY signal is allowed to populate both the signal and the control regions, and it is
scaled by a freely floating signal normalisation factor.
The background normalisation factors are also determined simultaneously in the fit.
A SUSY model with a specific set of sparticle masses is rejected if the upper limit at \percent{95} confidence level (CL)
of the signal normalisation factor obtained in this fit is smaller than unity.
\end{itemize}

The likelihood function is a product of the probability density functions, one for each region contributing to the fit.  
The number of events in a given CR or SR is described using a Poisson distribution, the mean of which is the sum of the expected contributions from all background and signal sources. 
The systematic uncertainties in the expected event yields are included
as nuisance parameters and are assumed to follow a Gaussian distribution with a width determined from the size of the uncertainty. 
Correlations between control and signal regions, and background processes are taken into account with common nuisance
parameters. The fit parameters are determined by maximising the product of the Poisson probability functions and the constraints for
the nuisance parameters.

\section{Systematic uncertainties}
\label{sec:Systematics}

Systematic uncertainties have an impact on the estimates of the background and signal event yields
in the control and signal regions. Uncertainties arising from experimental effects and theoretical sources are estimated.

The main sources of experimental systematic uncertainty in the SM
background estimates include tau lepton and jet energy calibrations and resolution, 
tau lepton identification, pile-up, and 
uncertainties related to the modelling of \met\ in the simulation. The uncertainties in the energy and momentum scale of each of the objects 
entering the \met\ calculation are estimated, as well as the uncertainties in the soft-term resolution and scale.
A variation in the pile-up reweighting of the MC simulated event samples is included to cover the uncertainty in the ratio of the predicted and 
measured inelastic cross section in the fiducial volume defined by $M_X >$ 13 GeV where $M_X$ is the mass of the hadronic system~\cite{Aaboud:2016mmw}.
The main contributions to experimental systematic uncertainties in the SR-lowMass (SR-highMass) are from the tau lepton identification and energy scale around \percent{6} (\percent{8}),
jet energy scale and resolution around \percent{11} (\percent{4}), \met\ soft-term resolution and scale around \percent{2} (\percent{6}), and pile-up around \percent{8} (\percent{8}).
Other contributions are less than \percent{3}.

Theoretical uncertainties affecting the MC event generator predictions are estimated by varying the renormalisation, factorisation, and resummation scales,
and the matching scale between the matrix elements and the parton shower. 
For $W$+jets and diboson processes, the uncertainties related to the
choice of the QCD renormalisation and factorisation scales are estimated from 
the comparison of the nominal samples with samples with these scales varied up and down by a factor of two. 
Uncertainties in the resummation scale and the matching scale between the matrix elements and parton shower are evaluated  by varying 
up and down the corresponding parameters in \SHERPA by a factor of two.
For $W$+jets events, the uncertainty due to the jet \pt\ threshold used for parton--jet matching 
is estimated by comparing the baseline samples with jet \pt\ threshold set to 20 GeV to samples with a threshold of 15 or 30 GeV. 
\SHERPA is compared with \MADGRAPH to estimate the uncertainty related to the generator choice for $W$+jets production.
The total theoretical uncertainty for diboson processes in the SRs is around \percent{15}, mainly coming from the choice of QCD renormalisation scale (4--\percent{9}) and resummation scale (around \percent{10}).
The theory uncertainty in $W$+jets production is 13--\percent{20}, and
the main source is the event generator uncertainty (\percent{4--17}) and the QCD renormalisation scale (\percent{9--10}).
An overall systematic uncertainty of \percent{6} in the inclusive cross section is assigned to the diboson process.
Based on previous studies~\cite{SUSY-2013-14}, a total theoretical uncertainty of \percent{25} is assigned for 
the top-quark and $Z$+jets contributions to the SRs.

The following sources of uncertainty are considered for the ABCD method used to 
determine the multi-jet background:  the correlation between the tau-id, the charge requirement, and the kinematic variables \mttwo, the
limited number of events in the CRs, and the subtraction of other SM backgrounds.
The systematic uncertainty in the correlation is estimated by comparing the transfer
factor from CR-B to CR-C to that of VR-E to VR-F. The systematic uncertainty in the non-multi-jet background subtraction in the control regions is estimated by considering the systematic
uncertainty of the MC estimates of the non-multi-jet background in the CRs. Both uncertainties are of the order of \percent{10}.
The systematic uncertainty in the signal region due to the limited number of events in the control regions is estimated by taking the statistical uncertainty of the event yields in these control regions.
It corresponds to the largest source of uncertainty for the ABCD method, and it reaches 21--\percent{42} for CR-A.

The systematic uncertainties on the background estimates in the SRs are summarised in Table \ref{tab:tauSR_syst}. 
The dominant uncertainties are the multi-jet background normalisation (around \percent{32} in both SR-lowMass and SR-highMass), 
and the statistical uncertainty of the MC predictions (around \percent{18} in SR-lowMass and \percent{24} in SR-highMass respectively). 

The total uncertainty in the signal yields for the SUSY reference points defined in Section~\ref{sec:data} is about \percent{20}. 
The main sources of experimental uncertainty are the tau lepton identification and energy scale, jet energy scale and resolution,
 \met\ soft-term resolution and scale, and pile-up: they amount to a
 total of about \percent{15}. The cross-section uncertainty is taken into account as main source of theoretical uncertainty, 
 and it varies from \percent{3} to \percent{20} for the considered SUSY models.
SUSY models with higher chargino mass have larger uncertainties.

\begin{table}[!hbp]   
\caption{ The relative systematic uncertainty (\%) in the background estimate in the SR-lowMass and SR-highMass from the leading sources. 
Uncertainties from different sources may be correlated, and do not necessarily add in quadrature to the total uncertainty.  
} 
\centering
\smallskip
 \smallskip
\begin{tabular}{|l|r|r|}
\hline
Source of systematic uncertainty & SR-lowMass & SR-highMass\\
\hline
\hline
Normalisation uncertainties of the multi-jet background & 32 & 32 \\
 Statistical uncertainty of MC samples & 18& 24\\
Multi-jet estimation & 14 & 13 \\ 
Pile-up reweighting & 8 & 8 \\
Jet energy scale and resolution & 11 & 4 \\
Tau identification and energy scale & 6& 8\\
$\met$ soft-term resolution and scale & 2 & 6\\
\hline
Total & 40 & 38 \\
\hline
\end{tabular}\label{tab:tauSR_syst}
\end{table}

\section{Results}
\label{sec:result}

\begin{figure}
\centering
\includegraphics[width=0.49\textwidth]{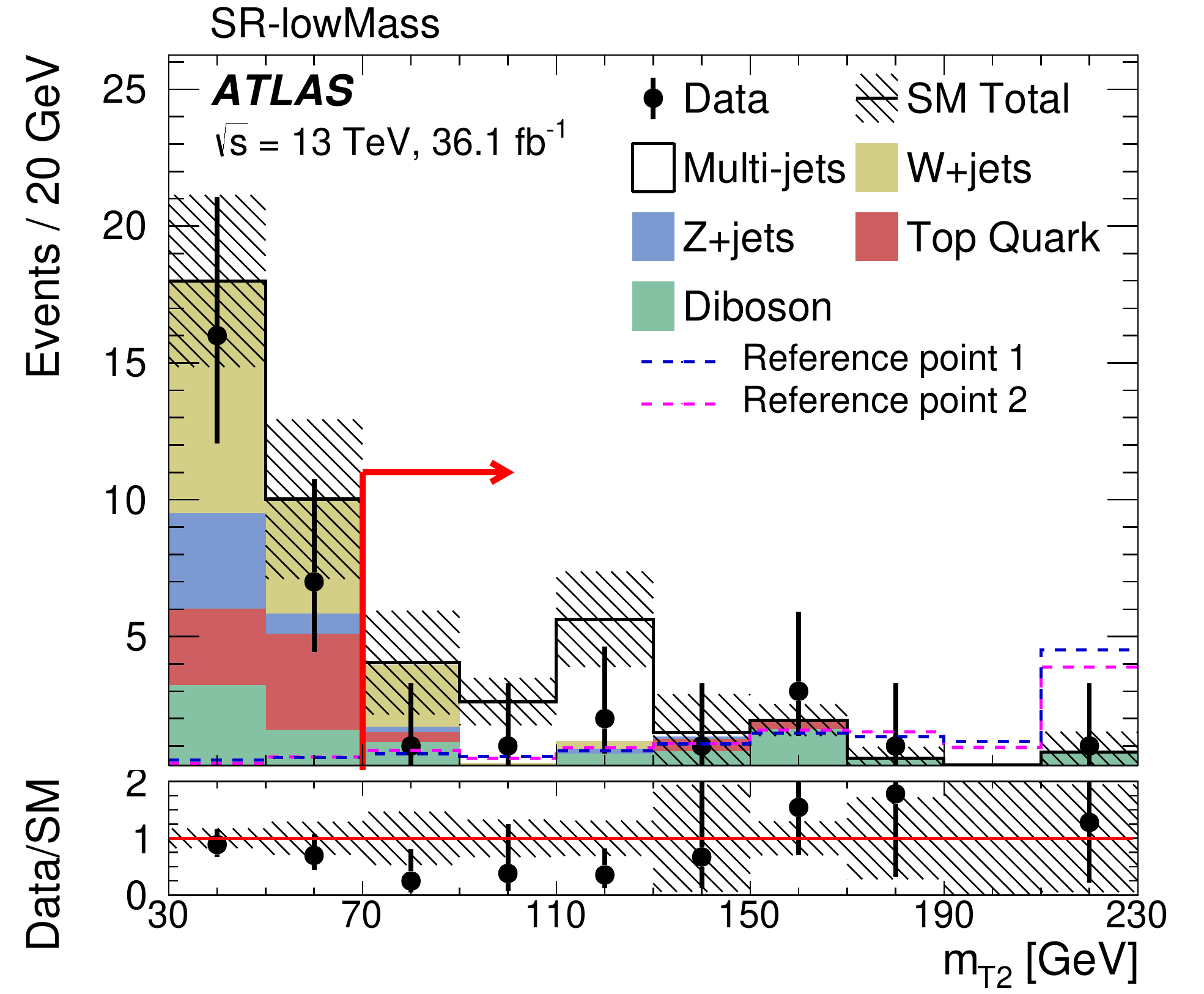}
\includegraphics[width=0.49\textwidth]{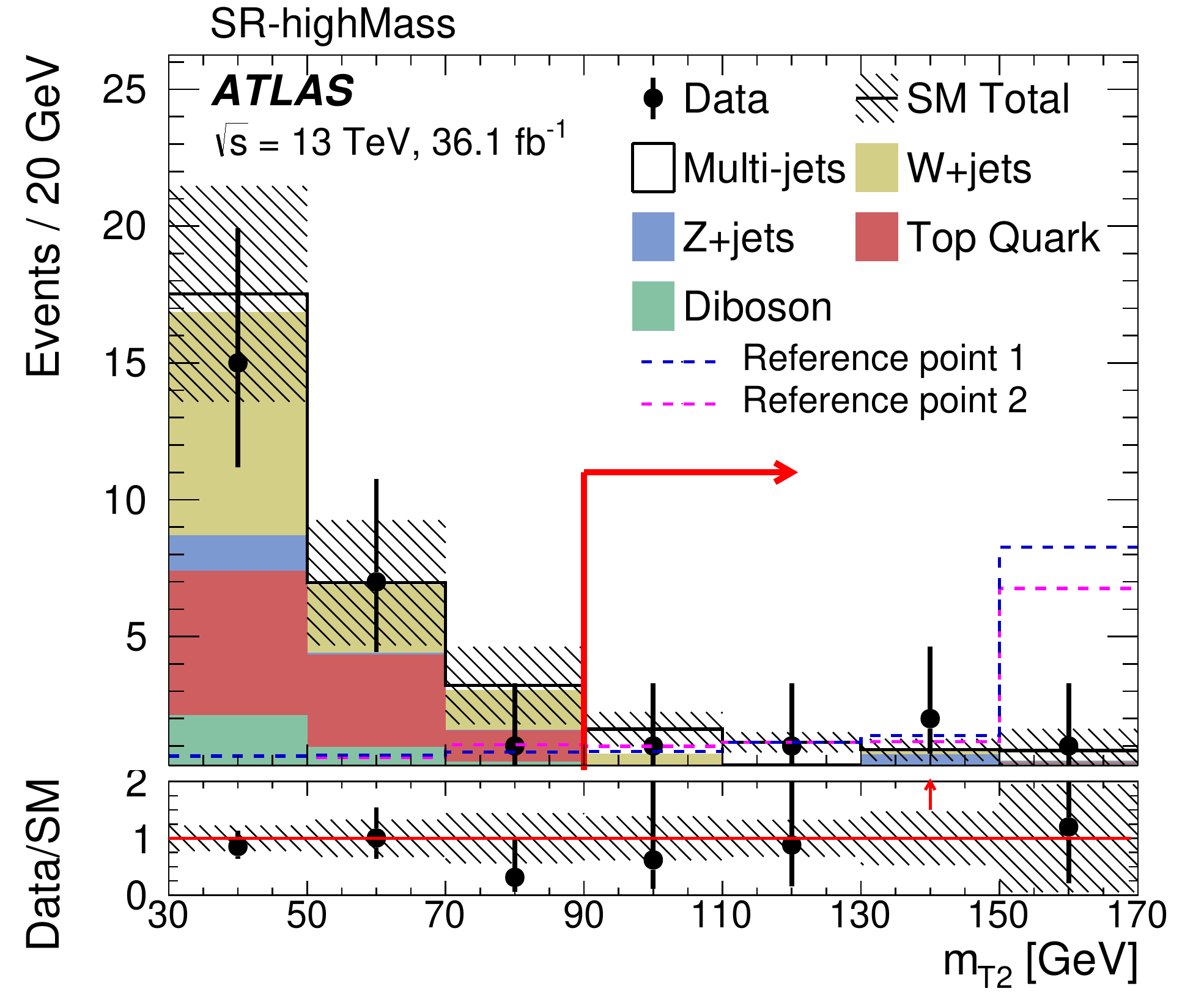}\\
\caption{The \mttwo\ distribution before the \mttwo\ requirement is
  applied for SR-lowMass (left) and SR-highMass (right) regions, where the arrow indicates the position of the cut in the signal region. 
The stacked histograms show the expected SM backgrounds.  
The multi-jet contribution is estimated from data using the ABCD method. 
The contributions of multi-jet and $W$+jets events are scaled with the
corresponding normalisation factors. The hatched bands represent the sum in quadrature of systematic and statistical uncertainties of the total SM background.
For illustration, the distributions of the SUSY reference points (defined in Section \ref{sec:data}) are
also shown as dashed lines. The lower panels show the ratio of data to
the total SM background estimate. The last bin includes the overflow events.
}
\label{fig:SR-result}
\end{figure}

The observed number of events in each signal region and the expected contributions from SM processes  
are given in Table~\ref{tab:results}. The contributions of multi-jet and $W$+jets events are scaled with the normalisation factors
obtained from the background-only fit described in Section~\ref{sec:fit}. 
The multi-jet normalisation with respect to the prediction from the ABCD method in the SR-lowMass (SR-highMass) is
compatible with unity and has an uncertainty of around \percent{100} (\percent{86}), due to 
the small number of observed events in the multi-jet CR-A.
The $W$+jets normalisation is 1.02$\pm$0.15. 
The \mttwo\ distribution is shown in Figure \ref{fig:SR-result} for data, expected SM backgrounds, and
the SUSY reference points defined in Section~\ref{sec:data}.
In both signal regions, observations and background predictions are
found to be compatible within uncertainties.

Upper limits at \percent{95} CL on the number of
non-SM events in the SRs are derived from the model-independent fit.
All limits are calculated using the \CLs prescription~\cite{CLs:2002}. Normalising these by the integrated luminosity of the data
sample, they can be interpreted as upper limits on the visible non-SM cross section, $\sigma_{\mathrm {vis}}^{95}$, which is
defined as the product of acceptance, reconstruction efficiency and production cross section. 
The accuracy of the limits obtained from the asymptotic formula was tested for all SRs by randomly generating a large number of pseudo-datasets and repeating the fit. Good agreement was found.

\begin{table}
\centering
\caption{ 
 Observed and expected numbers of events  in the signal regions.
The contributions of multi-jet and $W$+jets events are scaled with the normalisation factors
obtained from the background-only fit described in Section~\ref{sec:fit}. 
Expected event yields for the SUSY reference points (defined in Section~\ref{sec:data}) are also shown.
The uncertainties correspond to the sum in quadrature of statistical  and systematic uncertainties. 
The correlation of systematic uncertainties among control regions and
among background processes is fully taken into account.
The one-sided $p_0$-values, the observed and expected \percent{95} CL
upper limits on the visible non-SM cross section ($\sigma_{\mathrm
  {vis}}^{95}$) are given. Values of $p_0>0.5$  are truncated to $p_0=0.5$.
}
 \smallskip
 \smallskip
\begin{tabular}{|l|c|c|}
\hline
SM process  & SR-lowMass                            & SR-highMass    \\
\hline
\hline
Diboson        &  5.9 $\pm$  2.2 &  1.0 $\pm$  0.8    \\
$W$+jets      &  1.8 $\pm$  1.1 &  0.7 $\pm$  0.5  \\
Top quark     &  1.2 $\pm$  1.0 &  $0.03_{-0.03}^{+0.26}$ \\ 
$Z$+jets    &   $0.6_{-0.6}^{+0.7}$ &  0.6 $\pm$  0.5 \\
Multi-jet      &  4.3 $\pm$ 4.0  &  1.3 $\pm$ 1.1  \\
\hline
SM total            &  14 $\pm$  6\hd &  3.7 $\pm$  1.4    \\ 
\hline
Observed         &   10   &    5  \\
\hline     
Reference point 1  &   $11.8 \pm 2.8$\hd  & $11.6 \pm 2.6$\hd \\  
Reference point 2  &   $11.4 \pm 2.6$\hd  & $10.0 \pm 2.1$\hd \\  
\hline 
\rule{0pt}{2.6ex}
$p_0$                                                    &0.5    & 0.3         \\
Expected $\sigma_\text{vis}^{95}$ [fb]  & $ { 0.31 }^{ +0.12 }_{ -0.08 }$ &$ { 0.17 }^{ +0.08 }_{ -0.05 }$ \\ [+1mm]

Observed $\sigma_\text{vis}^{95}$ [fb]  & $0.26$  &    $0.20$   \\ 

\hline   
\end{tabular}
\label{tab:results}
\end{table}

\section{Interpretation}
\label{section:interpretation}

\begin{figure}
  \centering
  \includegraphics[width=0.49\textwidth]{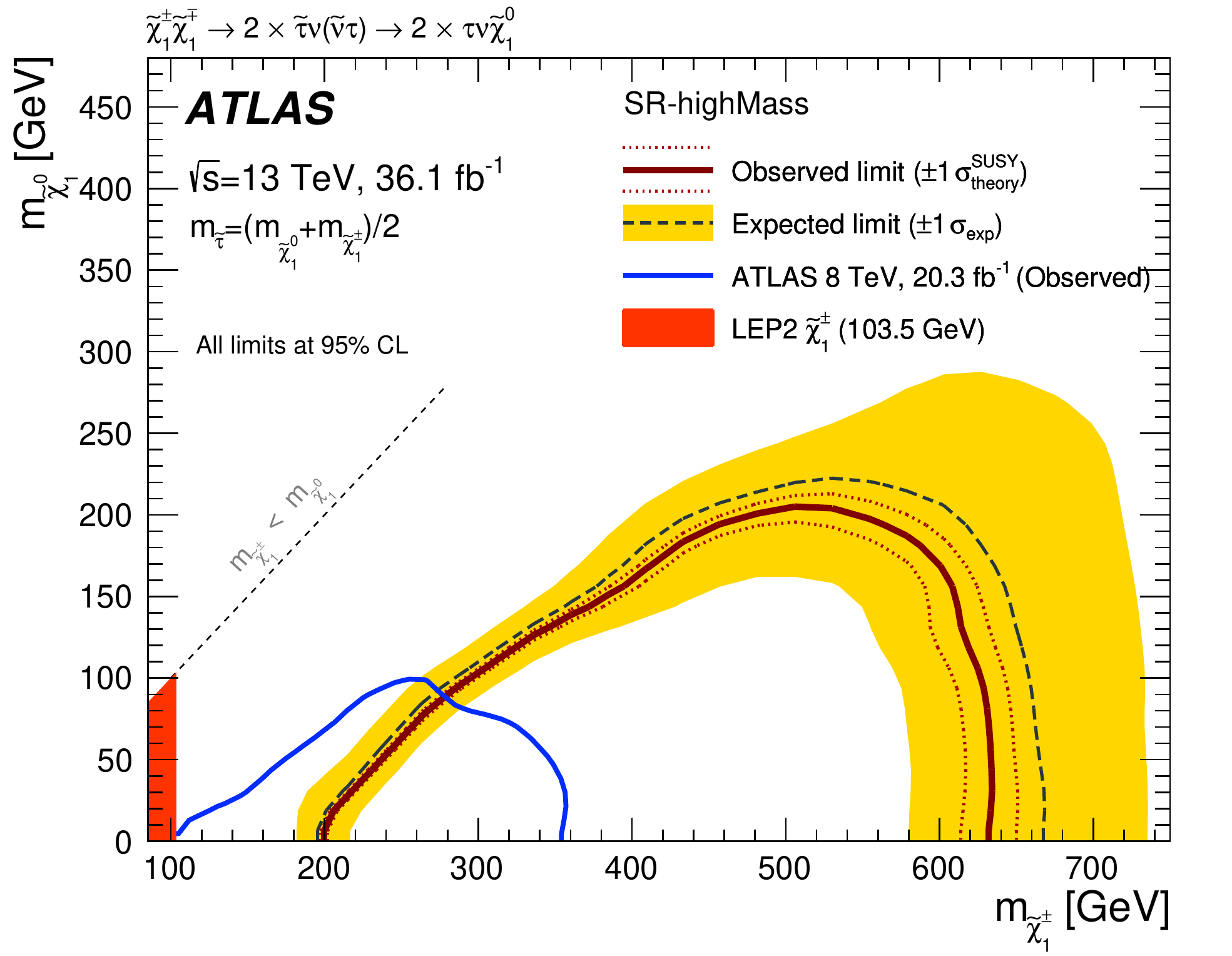}
  \includegraphics[width=0.49\textwidth]{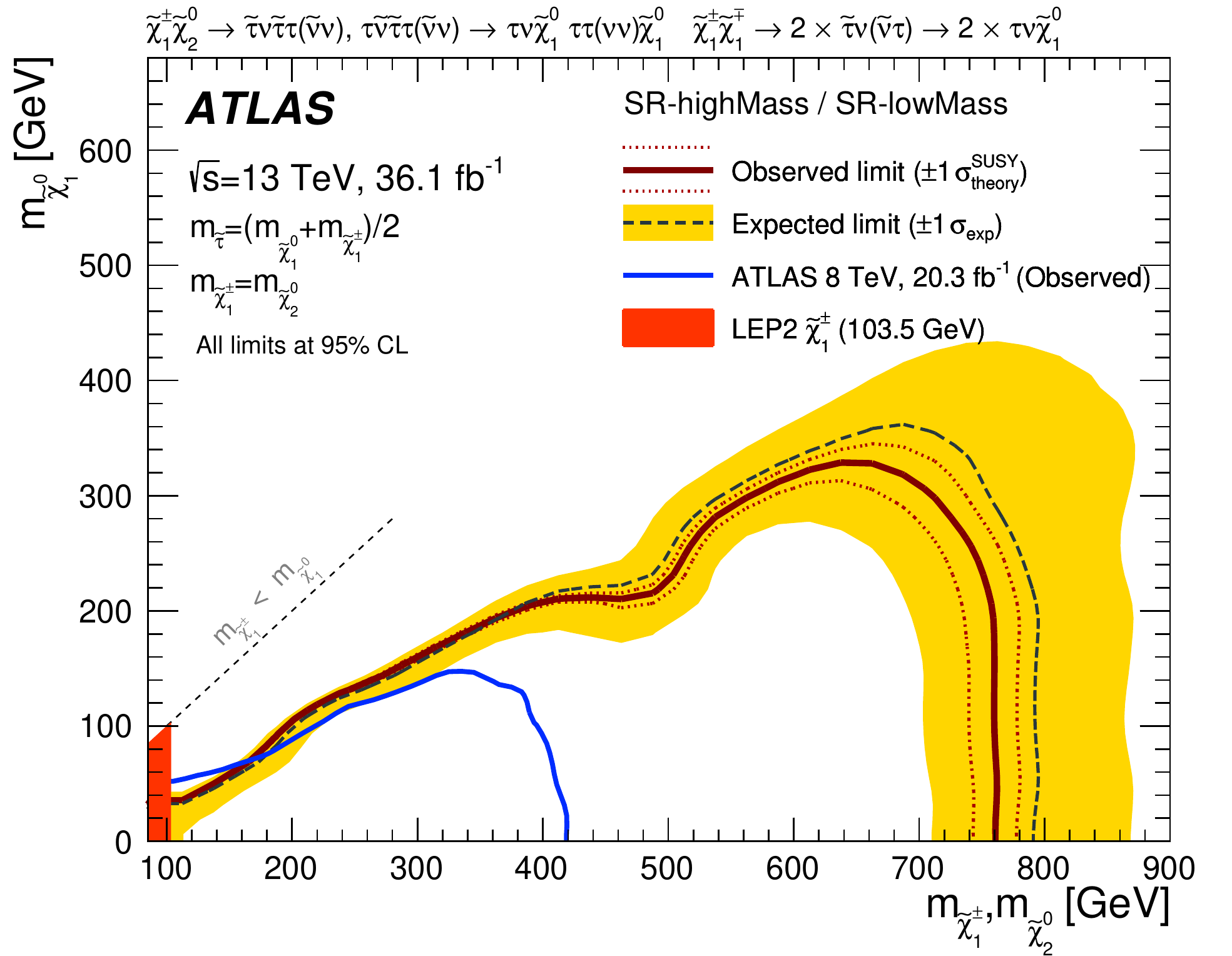}\\
  \caption{The \percent{95} CL exclusion contours for simplified models with \ConepConem\ production (left) and production 
  of \ConepConem\ and \ConepmNtwo (right).
The text provides details of exclusion curves and uncertainty bands.
The LEP limit on the chargino mass is also shown.
Results are compared with the observed limits obtained by
previous ATLAS searches \cite{SUSY-2013-14} as blue contours.
}
  \label{fig:alimit}
\end{figure}

In the absence of a significant excess over the expected SM background, the observed and expected numbers of events in the signal regions are used to
place exclusion limits at \percent{95} CL using the
model-dependent limit fit described in Section~\ref{sec:fit}. SR-highMass is used  to derive limits on \ConepConem\ production and
the best limit expected for SR-highMass and SR-lowMass is used to derive limits for the production of \ConepConem\ and \ConepmNtwo.
The exclusion limits for simplified models with $x=0.5$, described in Section \ref{sec:data}, 
are shown in Figure \ref{fig:alimit}. Only
\ConepConem\ production is assumed for the left plot,
whereas both production processes are considered simultaneously for
the right plot.
The solid (dashed) lines show the observed (expected) 
exclusion contours. The band around the expected limit shows the
$\pm1\sigma$ variations, including all
uncertainties except theoretical uncertainties in the signal cross section. 
The dotted lines around the observed limit indicate the sensitivity to
$\pm1\sigma$ variations of the theoretical uncertainties in the signal cross section. 
 
Chargino masses up to 630 GeV are excluded for a massless lightest neutralino in the scenario of direct production of chargino pairs.
In the case of production of chargino pairs and mass-degenerate charginos and next-to-lightest neutralinos, 
chargino masses up to 760 GeV are excluded for a massless lightest neutralino. Both limits apply to scenarios where 
the neutralinos and charginos decay solely via intermediate staus and tau sneutrinos, and with the parameter $x$ equal to 0.5.
These limits significantly extend previous results~\cite{SUSY-2013-14,CMS-SUS-13-006} in the high chargino mass
region. 

\begin{figure}
  \centering
  \includegraphics[width=0.49\textwidth]{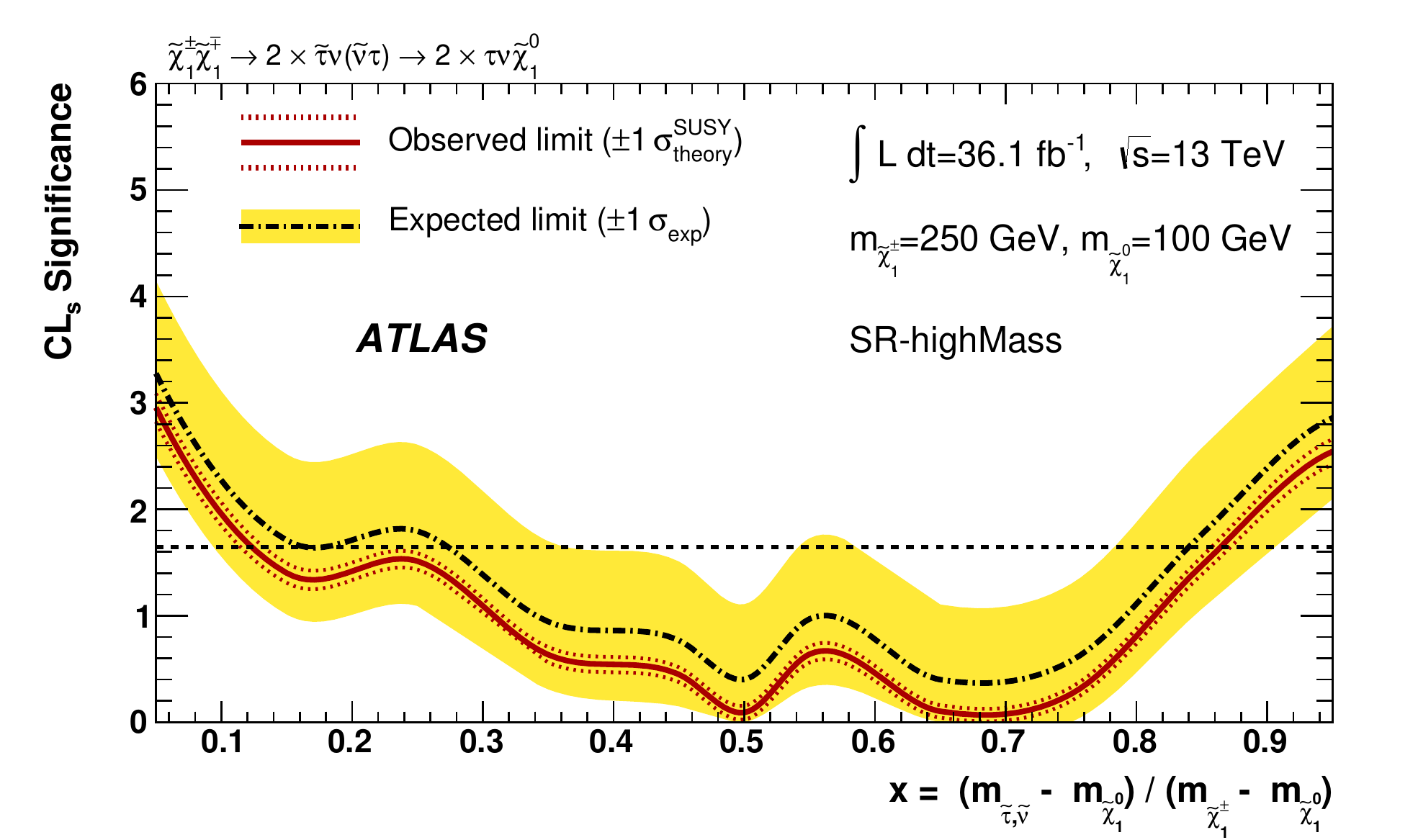}
  \includegraphics[width=0.49\textwidth]{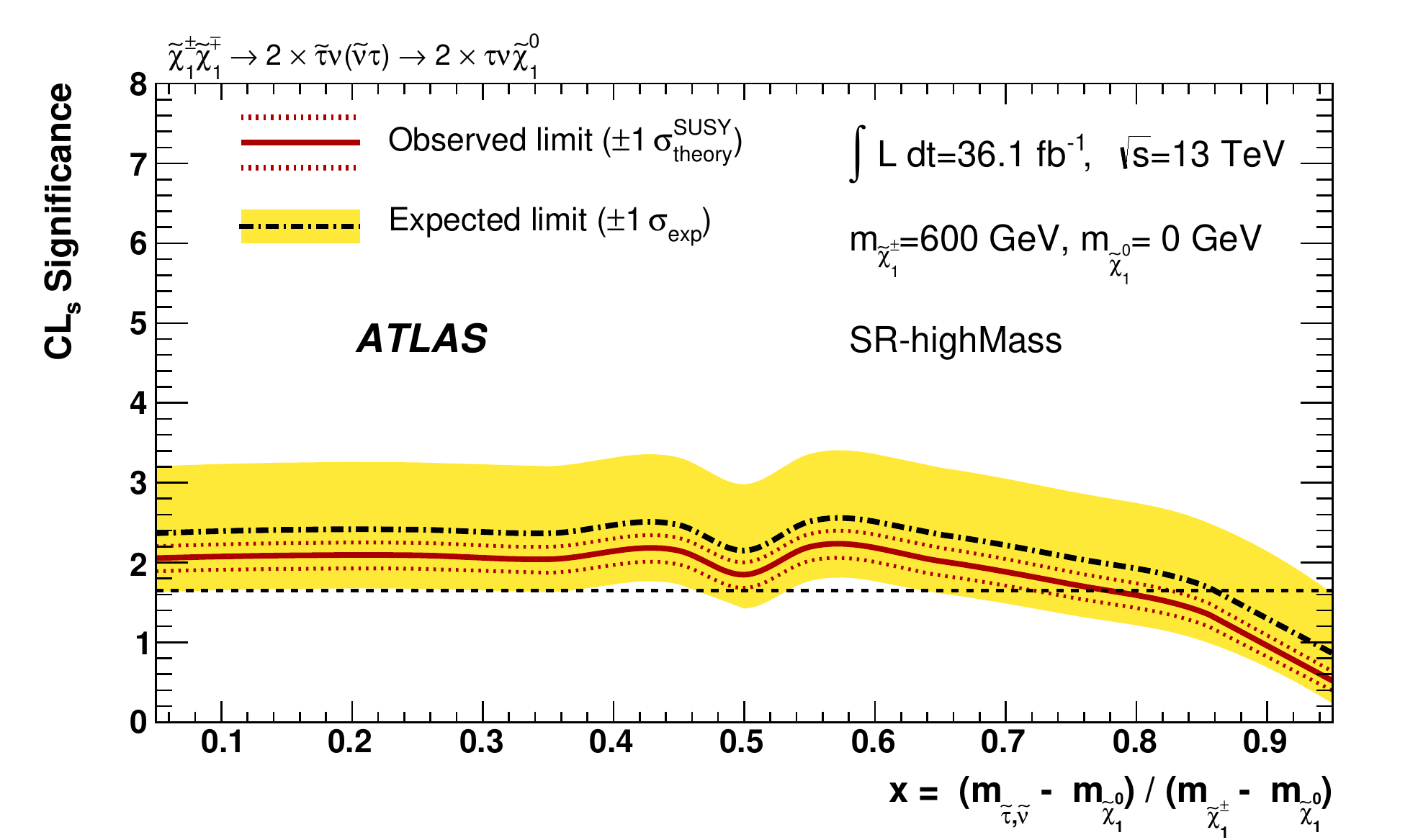}\\
  \includegraphics[width=0.49\textwidth]{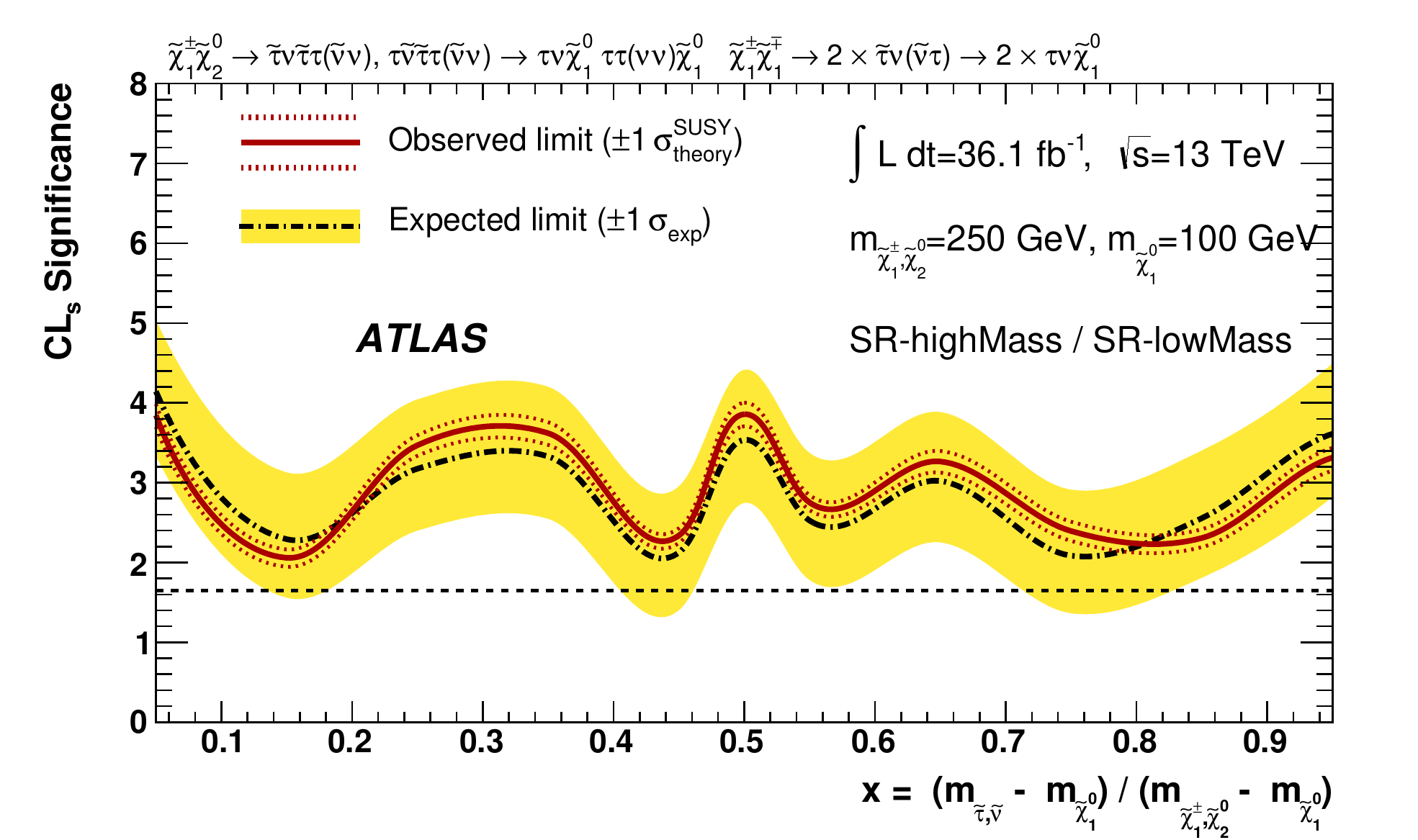}
  \includegraphics[width=0.49\textwidth]{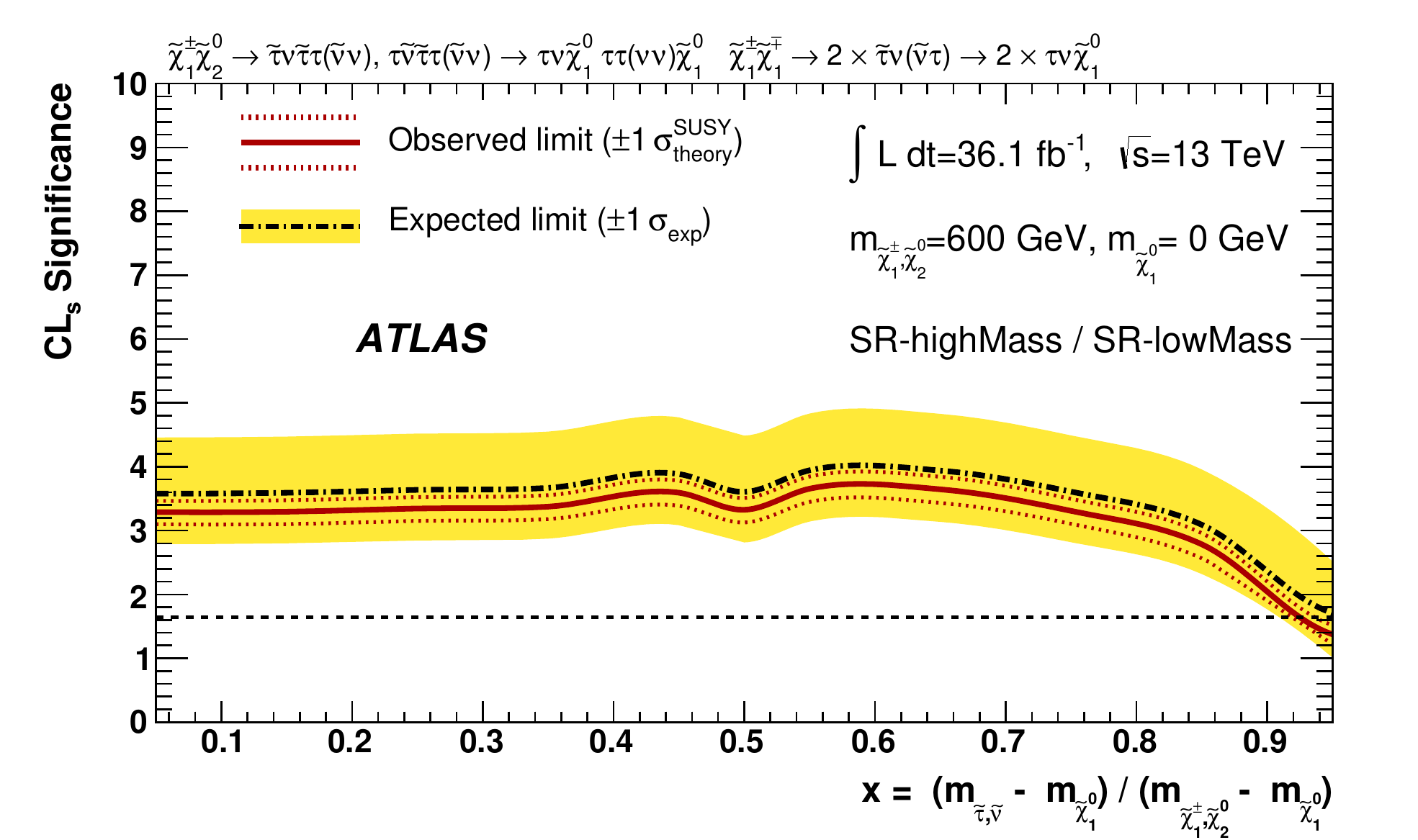}\\

  \caption{The \CLs significance for the benchmark models described in Section~\ref{sec:data} as a
function of the parameter $x$.
    The benchmark scenario with large mass splitting 
	($m(\chinoonepm,\ninotwo)$ = 600~GeV and $m(\ninoone)$ = 0~GeV) is shown on the right, 
	and the compressed benchmark scenario ($m(\chinoonepm,\ninotwo)$ = 250~GeV and $m(\ninoone)$ = 100~GeV) on the left, for \ConepConem\ production (top), and \ConepConem\ and \ConepmNtwo production (bottom). 
    SR-highMass is used for \ConepConem\ production, while the SR with the best expected \CLs value for each point of the parameter space is used for \ConepConem\ and \ConepmNtwo production.
    }
    \label{fig:variablex}
\end{figure}

\newcommand{\mttwox}{\mttwo\xspace}
The impact of $x$ different from 0.5 is studied by varying it between $0.05$ and $0.95$
for two benchmark scenarios. The \CLs significance as a function of the parameter $x$ is shown in Figure \ref{fig:variablex}.
When only \ConepConem\ production is considered, the benchmark scenario with large mass-splitting 
($m(\chinoonepm)$ = 600~GeV and massless \ninoone) can be excluded for $x$ up to $0.75$.
For larger values of $x$ the \pt spectra of the tau leptons from the chargino decay become very soft.
The compressed benchmark scenario ($m(\chinoonepm)$ = 250~GeV and $m(\ninoone)$ = 100~GeV)
can only be excluded for the extreme cases with $x=0.05$ or $x=0.95$, since the \mttwox requirement is
more effective for models with large mass-splittings between the charginos or the staus and the lightest neutralino. 
Models with low values of $x$ typically predict dark-matter relic density consistent with cosmological observations.
For combined production of \ConepConem\ and \ConepmNtwo the same general features are observed, 
but due to the higher signal yields with respect to \ConepConem\ production alone, 
both benchmark scenarios can be excluded for all considered values of $x$. 

\section{Conclusion}
 
Searches for the electroweak production of supersymmetric particles in events with
at least two hadronically decaying tau leptons are performed using \ourintlumi of
$pp$ collision data at $\sqrt{s}=13$~TeV recorded with the ATLAS experiment at the Large Hadron Collider.
Agreement between data and SM predictions is observed in two optimised
signal regions. The results are used to set limits on the visible cross section
for events beyond the Standard Model in each signal region. Observed upper limits on the simplified model cross-sections 
have been calculated and are available in~\cite{hepdata}. 

Exclusion limits are placed on parameters of simplified electroweak
supersymmetry models in scenarios where 
the neutralinos and charginos decay solely via intermediate left-handed staus and tau sneutrinos, and 
the mass of the $\tilde{\tau}_{\mathrm L}$ state is set to be halfway between the masses of the \chinoonepm\ and the \ninoone
($x=0.5$). 
Chargino masses up to 630 GeV are excluded for a massless lightest
neutralino in the scenario of direct production of 
chargino pairs, with each chargino decaying into the lightest neutralino via an intermediate on-shell stau or tau sneutrino. 
An additional benchmark scenario with large mass-splitting ($m(\chinoonepm)$ = 600~GeV and massless \ninoone) can be 
excluded for $x$ up to $0.75$, whereas a compressed benchmark scenario ($m(\chinoonepm)$ = 250~GeV and $m(\ninoone)$ = 100~GeV)
can only be excluded for the extreme cases with $x=0.05$ or $x=0.95$.
In the case of production of chargino pairs and mass-degenerate
charginos and next-to-lightest neutralinos, common \chinoonepm and \ninotwo\ masses up to 760 GeV are excluded for a massless lightest neutralino.
The additional benchmark scenarios with small and large mass-splitting can be both excluded for all considered values of $x$.

\section*{Acknowledgements}

We thank CERN for the very successful operation of the LHC, as well as the
support staff from our institutions without whom ATLAS could not be
operated efficiently.

We acknowledge the support of ANPCyT, Argentina; YerPhI, Armenia; ARC, Australia; BMWFW and FWF, Austria; ANAS, Azerbaijan; SSTC, Belarus; CNPq and FAPESP, Brazil; NSERC, NRC and CFI, Canada; CERN; CONICYT, Chile; CAS, MOST and NSFC, China; COLCIENCIAS, Colombia; MSMT CR, MPO CR and VSC CR, Czech Republic; DNRF and DNSRC, Denmark; IN2P3-CNRS, CEA-DRF/IRFU, France; SRNSF, Georgia; BMBF, HGF, and MPG, Germany; GSRT, Greece; RGC, Hong Kong SAR, China; ISF, I-CORE and Benoziyo Center, Israel; INFN, Italy; MEXT and JSPS, Japan; CNRST, Morocco; NWO, Netherlands; RCN, Norway; MNiSW and NCN, Poland; FCT, Portugal; MNE/IFA, Romania; MES of Russia and NRC KI, Russian Federation; JINR; MESTD, Serbia; MSSR, Slovakia; ARRS and MIZ\v{S}, Slovenia; DST/NRF, South Africa; MINECO, Spain; SRC and Wallenberg Foundation, Sweden; SERI, SNSF and Cantons of Bern and Geneva, Switzerland; MOST, Taiwan; TAEK, Turkey; STFC, United Kingdom; DOE and NSF, United States of America. In addition, individual groups and members have received support from BCKDF, the Canada Council, CANARIE, CRC, Compute Canada, FQRNT, and the Ontario Innovation Trust, Canada; EPLANET, ERC, ERDF, FP7, Horizon 2020 and Marie Sk{\l}odowska-Curie Actions, European Union; Investissements d'Avenir Labex and Idex, ANR, R{\'e}gion Auvergne and Fondation Partager le Savoir, France; DFG and AvH Foundation, Germany; Herakleitos, Thales and Aristeia programmes co-financed by EU-ESF and the Greek NSRF; BSF, GIF and Minerva, Israel; BRF, Norway; CERCA Programme Generalitat de Catalunya, Generalitat Valenciana, Spain; the Royal Society and Leverhulme Trust, United Kingdom.

The crucial computing support from all WLCG partners is acknowledged gratefully, in particular from CERN, the ATLAS Tier-1 facilities at TRIUMF (Canada), NDGF (Denmark, Norway, Sweden), CC-IN2P3 (France), KIT/GridKA (Germany), INFN-CNAF (Italy), NL-T1 (Netherlands), PIC (Spain), ASGC (Taiwan), RAL (UK) and BNL (USA), the Tier-2 facilities worldwide and large non-WLCG resource providers. Major contributors of computing resources are listed in Ref.~\cite{ATL-GEN-PUB-2016-002}.

\FloatBarrier
\clearpage

\clearpage
\bibliographystyle{bibtex/bst/atlasBibStyleWithTitle}
\bibliography{bibtex/bib/ATLAS,bibtex/bib/CMS,bibtex/bib/ConfNotes,bibtex/bib/PubNotes,susy,additional}



\onecolumn
\clearpage
\begin{flushleft}
{\Large The ATLAS Collaboration}

\bigskip

M.~Aaboud$^\textrm{\scriptsize 137d}$,
G.~Aad$^\textrm{\scriptsize 88}$,
B.~Abbott$^\textrm{\scriptsize 115}$,
O.~Abdinov$^\textrm{\scriptsize 12}$$^{,*}$,
B.~Abeloos$^\textrm{\scriptsize 119}$,
S.H.~Abidi$^\textrm{\scriptsize 161}$,
O.S.~AbouZeid$^\textrm{\scriptsize 139}$,
N.L.~Abraham$^\textrm{\scriptsize 151}$,
H.~Abramowicz$^\textrm{\scriptsize 155}$,
H.~Abreu$^\textrm{\scriptsize 154}$
% ...
\bigskip
\\
% ...
$^{12}$ Institute of Physics, Azerbaijan Academy of Sciences, Baku, Azerbaijan\\
% ...
$^{88}$ CPPM, Aix-Marseille Universit{\'e} and CNRS/IN2P3, Marseille, France\\
% ...
$^{115}$ Homer L. Dodge Department of Physics and Astronomy, University of Oklahoma, Norman OK, United States of America\\
% ...
$^{119}$ LAL, Univ. Paris-Sud, CNRS/IN2P3, Universit{\'e} Paris-Saclay, Orsay, France\\
% ...
$^{137}$ $^{(a)}$ Facult{\'e} des Sciences Ain Chock, R{\'e}seau Universitaire de Physique des Hautes Energies - Universit{\'e} Hassan II, Casablanca; $^{(b)}$ Centre National de l'Energie des Sciences Techniques Nucleaires, Rabat; $^{(c)}$ Facult{\'e} des Sciences Semlalia, Universit{\'e} Cadi Ayyad, LPHEA-Marrakech; $^{(d)}$ Facult{\'e} des Sciences, Universit{\'e} Mohamed Premier and LPTPM, Oujda; $^{(e)}$ Facult{\'e} des sciences, Universit{\'e} Mohammed V, Rabat, Morocco\\
% ...
$^{139}$ Santa Cruz Institute for Particle Physics, University of California Santa Cruz, Santa Cruz CA, United States of America\\
% ...
$^{151}$ Department of Physics and Astronomy, University of Sussex, Brighton, United Kingdom\\
% ...
$^{154}$ Department of Physics, Technion: Israel Institute of Technology, Haifa, Israel\\
$^{155}$ Raymond and Beverly Sackler School of Physics and Astronomy, Tel Aviv University, Tel Aviv, Israel\\
% ...
$^{161}$ Department of Physics, University of Toronto, Toronto ON, Canada\\
% ...
$^{a}$ Also at Department of Physics, King's College London, London, United Kingdom\\
$^{b}$ Also at Institute of Physics, Azerbaijan Academy of Sciences, Baku, Azerbaijan\\
$^{c}$ Also at Novosibirsk State University, Novosibirsk, Russia\\
$^{d}$ Also at TRIUMF, Vancouver BC, Canada\\
$^{e}$ Also at Department of Physics {\&} Astronomy, University of Louisville, Louisville, KY, United States of America\\
% ...
$^{*}$ Deceased
\end{flushleft}


\end{document}
